\documentclass[a4paper,12pt]{article}

\usepackage[utf8]{inputenc}
\usepackage{graphicx}        
\usepackage{color}

\def\mmm{\mathcal{V}}
\def\mmw{\mathcal{W}}

\def\lie{\mathcal{L}}

\def\pertp{\varepsilon}

\def\EP{E}

\def\gfam{\hat g}
\def\Supfam{\hat \Sigma}

\def\hfam{\hat h}
\def\kfam{\hat \kappa}

\def\gfamp{g}
\def\Supfamp{\Sigma}
\def\hfamp{h}
\def\kfamp{\kappa}

\def\gback{g}
\def\fpt{K^{(1)}{}}
\def\spt{K^{(2)}{}}

\def\axial{\eta}
\def\stat{\xi}

\def \tp {t_+}
\def \phip {\varphi_+}
\def \rrp {r_+}
\def \thetap {\theta_+}
\def \tm {t_-}
\def \phim {\varphi_-}
\def \rrm {r_-}
\def \thetam {\theta_-}

\def \om {\omega^-}

\def\ro{a} 

\def\OH{\Omega^H}

\def\hH{h^H}
\def\mH{m^H}
\def\kH{k^H}

\def\oppert{\omega'}
\def\topert{(\omega-\Omega)}

\def\defi{:=}
\def\Qtwo{Q_2}
\def\Qt2{\hat{Q}_2}

\def\Qtwoz{\Qtwo{}_{(0)}}
\def\Qtwot{\Qtwo{}_{(2)}}

\def\Qttwoz{\Qt2{}_{(0)}}
\def\Qttwot{\Qt2{}_{(2)}}

\def\defor{\Xi}

\def\energy{E}
\def\pressure{P}
\def\Eb{\energy}
\def\Ep{\energy^{(1)}}
\def\Epp{\energy^{(2)}}
\def\Eppz{\energy^{(2)}_0}
\def\Eppt{\energy^{(2)}_2}
\def\Pb{\pressure}
\def\Pp{\pressure^{(1)}}
\def\Ppp{\pressure^{(2)}}
\def\Pppz{\pressure^{(2)}_0}
\def\Pppt{\pressure^{(2)}_2}

\def\pressurestar{\mathcal{P}}

\def\Pspp{\pressurestar}

\def\tpressurestar{\tilde\mathcal{P}}
\def\Ptspp{\tpressurestar}
\def\Ptsppz{\Ptspp_0}
\def\Ptsppt{\Ptspp_2}

\def \Hzero {\tilde{h}_0}
\def \Mzero {\tilde{m}_0}

\def \Htwo {\tilde{h}_2}
\def \Mtwo {\tilde{m}_2}
\def \Ktwo {\tilde{k}_2}

\def \fcross {f}
\def \fr {\fcross_2}

\def \kk {k_0}

\def \hk {h_0}

\def \mk {m_0}

\def \kkt {k_2}

\def \hkt {h_2}

\def \mkt {m_2}

\def \difk {\left[k\right]}
\def \difkz {\left[k_0\right]}
\def \difkt {\left[k_2\right]}

\def \difkp {\left[k'\right]}
\def \difkpz {\left[k'_0\right]}
\def \difkpt {\left[k'_2\right]}

\def \difm {\left[m\right]}
\def \difmz {\left[m_0\right]}
\def \difmt {\left[m_2\right]}

\def \difh {\left[h\right]}
\def \difhz {\left[h_0\right]}
\def \difht {\left[h_2\right]}

\def \difhp {\left[h'\right]}
\def \difhpz {\left[h'_0\right]}
\def \difhpt {\left[h'_2\right]}

\def \difg {\left[f\right]}
\def \diffr {\left[\fr\right]}

\def\qper{{\hfamp^{(1)}}}
\def\qperper{{\hfamp^{(2)}}}

\def\Tone{{T_1}}
\def\Ttwo{{T_2}}

\def\Qone{Q_1}
\def\Qtwo{Q_2}

\def\Kper{\fpt}
\def\Kperper{\spt}

\def\kappaper{{\kfamp^{(1)}}}
\def\kappaperper{{\kfamp^{(2)}}}

\def\Kpernornor{{Y^{}}}
\def\Kperpernornor{Y^{(2)}}
\def\Kpertan{{\Kper^{\,t}}}
\def\Kpernortan{{\tau^{}}}

\newcommand{\Sper}{{{\cal S}^{(1)}}}
\newcommand{\Sperper}{{{\cal S}^{(2)}}}
\def\grad{\mathrm{\scriptscriptstyle grad}}

\newtheorem{theorem}{Theorem}
\newtheorem{proposition}{Proposition}

\def\fin{\hfill \rule{2.5mm}{2.5mm}\\ \vspace{0mm}}
\def\finn{\hfill \rule{2.5mm}{2.5mm}}

\begin{document}
\title{Revisiting Hartle's model using perturbed matching theory to second order:
amending the change in mass}

\author{
Borja Reina and Ra\"ul Vera\\
Dept. of Theoretical Physics and History of Science,\\ University of the Basque Country UPV/EHU,\\
644 PK, Bilbao 48080, Basque Country, Spain}
\date{}
\maketitle

\begin{abstract}
Hartle's model describes the equilibrium configuration of a rotating
isolated compact body in perturbation theory up to second order in General Relativity.
The interior of the body is a perfect fluid with a barotropic equation of state,
no convective motions and rigid rotation. That interior is matched across
its surface to an asymptotically flat vacuum exterior.
Perturbations are taken to second order around a static and spherically symmetric
background configuration.
Apart from the explicit assumptions,
the perturbed configuration is constructed upon some implicit premises,
in particular the continuity of the functions describing the perturbation
in terms of some background radial coordinate.
In this work we revisit the model within
a modern general and consistent theory of perturbative matchings to second order,
which is independent of the
coordinates and gauges used to describe the two regions to be joined.
We explore the matching conditions up to second order in full.
The main particular result we present is that
the radial function $m_0$ (in the setting of the original work)
of the second order perturbation tensor, contrary to the original assumption,
presents a jump at the surface of the star, which is proportional
to the value of the energy density of the background
configuration there. As a consequence, the change in mass $\delta M$
needed by the perturbed configuration to keep the value of the central energy density
unchanged must be amended. 
We also discuss some subtleties that arise when studying the deformation of the star.

\end{abstract}

\section{Introduction}
Hartle's model \cite{Hartle1967} constitutes the basis of most of the analytical studies
performed to study slowly rotating stars in General Relativity (GR).
The formalism provides a method to construct numerical schemes in
axial symmetry \cite{LivRev_Stergioulas}.
The model describes the axially symmetric equilibrium configuration
of a rotating isolated compact body and its vacuum exterior in perturbation theory in GR.
The interior of the body is a perfect fluid which satisfies a
barotropic equation of state, does not have convective motions and rotates rigidly.
This is matched to a stationary and axisymmetric asymptotically flat vacuum
exterior region across a timelike hypersurface, and the whole model is assumed to have
equatorial symmetry. By matching we mean that there is no shell of matter on the surface of the star.
The approach is analytic, and makes use of a perturbative method
for slow rotation around a spherically symmetric static configuration driven by a single
parameter $\OH$\footnote{In order to ease the comparison with the original paper
\cite{Hartle1967} we will use a superscript $^H$ to indicate that
any object $f^H$ here refers to $f$ in \cite{Hartle1967}.}

The first order perturbation, driven by a single function $\omega^H$, accounts
for the rotational dragging of inertial frames. It does not change the shape of the
surface of the star. 
The second order perturbation, in contrast, does affect the original
spherical shape of the body, in agreement with the fact that
this must be independent of the sense of rotation.
The second order perturbation of the metric is described by three functions,
$h^H$, $m^H$ and $k^H$. In addition to the deformation of
the star, these functions provide the relation between the central density of the star, which
is kept unperturbed, and the excess of mass $\delta M$ between the perturbed and the static
background configuration needed to keep the central density of the star unchanged,
in analogy to the Newtonian approach (see \cite{Chandra_Poly_Newton_1933,ChandraLebo_Newton_1962}).

Apart from the explicit assumptions made in devising the model, the
construction of the perturbed configuration hides some seemingly important implicit assumptions.
In this paper we focus on one of those implicit assumptions, namely
the fact that the (perturbed) metric is written globally in terms of a single set of
spherical-like coordinates $\{t,r,\theta,\phi\}$,
that cover both the interior region (star) and exterior vacuum ($r\in (0,\infty)$),
in which the function $\omega^H$ is differentiable
and $h^H$, $m^H$ and $k^H$ are continuous. Explicitly,
\begin{eqnarray}
ds^2 &=& -e^{\nu(r)}\left(1+2  h^H(r,\theta)\right)dt^2
+ e^{\lambda(r)}\left(1+ 2 m^H(r,\theta)/(r-2M) \right)dr^2 \nonumber \\
&& + r^2(1+2  k^H(r,\theta))\left[d\theta^2 + \sin ^2 \theta (d\varphi- \omega^H(r,\theta) dt)^2\right]\,
+ \mathcal{O}((\OH)^3). \label{metricHartle}
\end{eqnarray}
Furthermore, the radial coordinate $r$ is fixed by imposing that
the function $k^H$ has no $l=0$ term in a Legendre expansion,
that is $k^H=k^H_2(r)P_2(\theta)+\ldots$. We will refer to that choice as the $k$-gauge.

In the theory of matching of spacetimes, in the exact case,
the existence of (Lichnerowicz) admissible coordinates, for which the metric functions are of class $C^1$,
once the matching of spacetimes is performed is known (c.f. \cite{Bonnor_Vickers,Mars99}).
However, how this fact translates to a perturbative scheme remains to be settled.
That is, the whole background configuration (interior and exterior) can indeed be described
by a metric with $C^1$ functions, but the differentiability (and even continuity) of
the functions describing the perturbations in some convenient gauge is not ensured a priori.
In any case,
a priori explicit choice of coordinates in which the metric and its perturbations
satisfy certain continuity
and differentiability conditions may constitute an implicit
assumption that, in principle at least, could subtract generality to the model.
More dramatically, it could turn out to be a wrong choice, and lead to wrong outcomes.

To analyse with rigour the consequences the choices of coordinates may have in Hartle's model, we
present here the study of the matching problem by making use of the perturbed matching theory, only
achieved in full generality and to second order in \cite{Mars2005}.
The global model is separated as the interior and exterior perturbed problems,
matched across a perturbed matching hypersurface. The scheme we present is
independent of any choice of coordinates at either side.
The two problems are described in terms of two sets of functions, as described above,
one for the interior and another for the exterior. To ease the reading we say a function 
is continuous if the values of the function at one point at one
side of the matching hypersurface 
agree with the values of the corresponding function 
computed at the corresponding (same) point at the other side.

The main result we prove is that in the initial coordinates (\ref{metricHartle})
used in the original work \cite{Hartle1967},
the function $\omega^H$ and its radial derivative 
can be taken to be continuous (see also \cite{ReinaVera_ERE2012}), and that
in the $k$-gauge, the functions $h^H$ and $k^H$ are continuous, in agreement with
\cite{Hartle1967}. However, the $l=0$ sector of the function $m^H$, $m^H_0(r)$, is not continuous
in general,
contrary to the implicit assumption in \cite{Hartle1967}. The discontinuity of $m^H_0(r)$ turns out
to be proportional to the value of the energy density of the background static
and spherically symmetric configuration 
at the surface of the star.
The consequence of this jump is that the calculation of the change in mass $\delta M$
of the perturbed star needed to keep the central density of the star unchanged
must be readdressed.
The amended expression in terms of the functions used in \cite{Hartle1967}
is given by equation (\ref{eq:deltaM}).

The importance of an amended expression for $\delta M$ lies, of course, in the existence
of cases for which that correction is relevant. Indeed, for any models, i.e. equations
of state, in which the energy density vanishes at the boundary of the star
the expression for $\delta M$ in \cite{Hartle1967} provides the correct value. This includes
the models studied in all the subsequent works by Hartle and Thorne,
in particular in \cite{HartleThorne1968} and \cite{HartleThorne1969}.
As far as we know, most of the candidate models for neutron stars satisfy indeed that condition. However,
some equations of state suitable to describe  strange quark matter stars yield
a non-zero energy density at the boundary of the star (see e.g. \cite{ColpiMiller, CuchiLinear, Lattimer2012}).
At least in that case the correcting factor in $\delta M$ could be of (numeric) relevance.
Another particular and simple case corresponds to stars with a constant energy density,
as those studied in \cite{Chandra_Miller1974}.
The correction to $\delta M$ properly applies in those cases, and a numerical analysis shows that 
the portion of the correction over the mass computed with the missing term is not negligible 
by any means \cite{Borja_constant_rho}.
The fact that some relevant second order function may had a jump across the boundary has
appeared previously
in the literature, explicitly in e.g. \cite{Bradley_etal2007} (see equation (46) there)
and in \cite{Tesis_D_Eriksson},
where a correct expression of $\delta M$ is given.
It is also implicit in other works such as
\cite{Gonzalez_Romero_etal}. 
However, the exact relationship of the functions
with the original $m^H_0(r)$ and thus the discrepancy in the computation of $\delta M$ in \cite{Hartle1967},
had not been realised at the moment. 
On the other hand,  let us stress that
the starting point of the
matching procedures using perturbative schemes found so far in the literature 
has always relied on the prescription of the matching hypersurface
defined as the set of points where the (perturbed) pressure vanishes (in a certain gauge).
However, and although physically reasonable,
that should be found as a consequence of the matching procedure, which ought to be
constructed from first principles only.

Let us remark that the present paper is one of a series of works aimed
at revisiting Hartle's model within the modern perturbation theory,
and perturbative matchings theory in particular, and thus put the model on firm grounds.
In the present paper we aim specifically to the (perturbed) matching conditions.
Some aspects which are shown in \cite{Hartle1967} by using the continuity
of functions, mainly the structure of the metric functions (no $l>2$ sector),
are still to be proven given the present state of things. 
That work, done in parallel, started in \cite{ReinaVera_ERE2012}
by using the framework construted in \cite{MaMaVe2007}, consisting
of a completely general
perturbative approach to second order around static configurations of the exterior
(asymptotically flat) vacuum problem of stationary and axisymmetric bodies with
arbitrary matter content. 
All in all, given that the matching conditions may be needed in more general situations
in future works, we have preferred to keep some generality in the results
by focusing first on a purely geometric setting in order to impose the field equations later.

The paper is structured as follows. In Section \ref{sec:overview} we first briefly review and present
the theory of perturbed matchings to second order from \cite{Mars2005}. That is followed by
the set up of the perturbed schemes needed for the (stationary and axially symmetric)
geometries that are going to be used
for the interior and exterior regions, together with the perturbed matching hypersurface.
We present, in the form of two propositions, the necessary and sufficient
conditions that the first and second order perturbations of the geometries at either side
and the perturbed hypersurface must satisfy in order to match.
Up to this point no Einstein's field equations have been imposed and thus the results are purely geometric.
In Section \ref{sec:pf-vacuum} the interior and exterior problems at first and second order
are imposed using  Hartle's model explicit assumptions: rigidly rotating perfect fluid interior
with barotropic equation of state, asymptotically flat vacuum exterior, and global equatorial symmetry.
The particularizaton of the previous propositions to Hartle's setting
is then analysed in detail. The result concerning the interior and exterior
problems at second order is finally given in the form of a Theorem,
in which the equations that the functions at either side must satisfy together
with their corresponding matching conditions are given in full.
Up to this point the whole problem has been treated within a somewhat general family of spacetime
gauges, that includes the two gauges employed in \cite{Hartle1967}
and most commonly used in the literature, namely the $k$-gauge described above
and the gauge that follows the surfaces of constant energy density (or pressure).
In Section \ref{sec:conclusion} we conclude by making explicit the link between the results
shown in the preceeding sections with
the results as presented in \cite{Hartle1967}.
The discontinuity of the $m^H_0(r)$ function is thus shown, and the correct
expression for $\delta M$ in terms of the functions used in \cite{Hartle1967} is given.

We devote an Appendix to discuss the deformation of the shape of the star, and show
how the description of the perturbed hypersurface in terms of the vanishing
of the pressure (in the gauge that follows surfaces of constant energy density, or pressure)
holds true in the end.
Let us note that the discontinuity found in the function $m^H_0(r)$ does not affect
the deformation.

In this paper we use $G=c=1$, greek indices for spacetime objects
and latin indices for objects relative to hypersurfaces.
Spacetimes with boundary (before matching) are assumed to be $C^3$
(as manifolds) and oriented, with oriented boundary if any.

\section{Perturbed matching to second order in brief}

\label{sec:overview}

We take the view of modern spacetime and matching perturbation theory.
A convenient starting notion in spacetime perturbation theory is a one-parameter
family of spacetimes $\{(\mmm_{\pertp},\gfam_\pertp)\}$ with diffeomorphically related manifolds,
from where we single out a background spacetime say $(\mmm_0,\gback)$
with $\mmm_0\defi \mmm_{\pertp=0}$ and $\gback\defi \gfam_{\pertp=0}$.
The points at each manifold of the family are
identified through a diffeomorphism, say $\psi_\pertp:\mmm_0\to\mmm_\pertp$.
This allows us to pull back $\gfam_\pertp$ onto the background spacetime,
and thus define a family of tensors $\gfamp_\pertp\defi \psi^*_\pertp(\gfam_\pertp)$
on $(\mmm_0,g)$, where $\gback\defi \gfam_{\pertp=0}=\gfamp_{\pertp=0} $,
a single manifold.
The metric perturbation tensors  are simply defined as the derivatives  of $\gfamp_\pertp$
with respect to $\pertp$ evaluated on $\pertp=0$ at each order of derivation.
$\fpt$ and $\spt$ will refer to the first and second
metric perturbation tensors.
At this point matter fields are also introduced as a $\pertp$-family of energy-momentum tensors $T_\pertp$
on  $(\mmm_0,g)$, and the corresponding perturbations are defined again by taking $\pertp$-derivatives.
Spacetime perturbation theory then consists of the study of the tensor fields $\fpt$ and $\spt$
satisfying certain field equations on a fixed background $(\mmm_0,g)$.

Spacetime perturbation theory carries, by construction, an inherent freedom, which lies precisely
on the freedom in choosing the diffeomorphism $\psi_\pertp$
identifying points of the manifolds. This is the so-called (spacetime) gauge freedom.
Different choices of identifications lead to different, but geometrically equivalent,
metric perturbation tensors. At each order in the perturbation a change of gauge
is described by a vector field on the background, which measures the shift between
identifications at each order. More explicitly, a change of gauge defines a $\pertp$-parameter
diffeomorphism, say $\Omega_\pertp:\mmm_0\to\mmm_0$. The first and second order gauge vectors,
denoted as $\vec s_1$ and $\vec s_2$,
can then be defined as \cite{Mars2005}
\begin{equation}
\vec s_1\defi\partial_\pertp \Omega_\pertp|_{\pertp=0}\qquad
\vec V_2\defi \partial_\pertp(\partial_h(\Omega_{h+\pertp}\circ\Omega^{-1}_\pertp)|_{h=0})|_{\pertp=0},\qquad
\vec s_2\defi \vec V_2+\nabla_{\vec s_1} \vec s_1.
\label{eq:s_gauges}
\end{equation}
Indicating with a ${}^{g}$ superscript a ``gauge transformed'' quantity,
the metric perturbation tensors thus transform as
\cite{Bruni_et_al_1997} (see \cite{Mars2005})
\begin{eqnarray}
      &&\fpt^{g}{}_{\alpha\beta}=\fpt{}_{\alpha\beta} + \lie_{\vec s_1} g{}_{\alpha\beta},\nonumber\\
      &&\spt^{g}{}_{\alpha\beta}=\spt{}_{\alpha\beta}
      +\lie_{\vec s_2} g_{\alpha\beta}+
      2\lie_{\vec s_1}\spt{}_{\alpha\beta} 
      -2s_1^\mu s_1^\nu R_{\alpha\mu\beta\nu} + 2\nabla_\alpha s_1^\mu \nabla_\beta s_{1\mu}.
\label{eq:Ksgauged}
\end{eqnarray}

The matching of two spacetimes with boundary, say $(\mmm^+,g^+,\Sigma^+)$ and $(\mmm^-,g^-,\Sigma^-)$,
requires an identification of the boundaries, $\Sigma^+$ and $\Sigma^-$.
If the boundaries are nowhere null (non-degenerate) the matching conditions (in full, so that
the global Riemann tensor shows no Dirac-delta term) demand the
equality of their respective first $h$ and second $\kappa$
fundamental forms.
The identification of the boundaries allows the construction of
an abstract manifold $\Sigma$ on which the first and second fundamental forms
as coming from both sides,
$\hfamp^\pm$ and $\kfamp^\pm$, are pulled back so that they can be compared.
The matching conditions demand the existence of one such identification
for which the first and second fundamental forms agree. In particular,
$\Sigma$ is endowed with the metric $\hfamp(=\hfamp^+=\hfamp^-)$.

To study perturbation theory on a background spacetime constructed from
the matching of two spacetimes one can use again the same picture.
We assume two families of spacetimes with boundary \footnote{We refer to \cite{Mars2005}
for a proper discussion on the subtleties involved in the definition of families
of spacetimes with boundary. Also, we need only to consider non-degenerate hypersurfaces $\Supfam_\pertp$,
without loss of generality. Their orientation will extend through $\pertp$ by continuity.}
$\{(\mmm^\pm_\pertp,\gfam^\pm_\pertp,\Supfam^\pm_\pertp)\}$
are matched across their respective boundaries $\Supfam^\pm_\pertp$ for each $\pertp$,
so that there exists a corresponding family of
diffeomorphically related hypersurfaces $\Supfam_\pertp$
on which 
the first and second fundamental forms from each side are equated,
$\hfam^+_\pertp=\hfam^-_\pertp$, $\kfam^+_\pertp=\kfam^-_\pertp$.
The matching hypersurface of the background configuration is $(\Supfamp_0,\hfamp)$,
where $\Supfamp_0\equiv \Supfam_0$ and $\hfamp= \hfam^+_0=\hfam^-_0$.
The idea is to construct, from those tensors on $\Supfam_\pertp$, corresponding families
 $\hfamp^\pm_\pertp$ and $\kfamp^\pm_\pertp$ on $(\Sigma_0,\hfamp)$ containing
also the information about how $\Sigma_0^\pm$ are perturbed
with respect to the gauges defined at each side $\psi^\pm_\pertp$,
which we want to keep free.
Taking $\pertp$-derivatives on $\pertp=0$ one can thus
construct $\hfamp^{(1)}$, $\hfamp^{(2)}$, $\kfamp^{(1)}$ and $\kfamp^{(2)}$
at first and second order at each side.
The matching conditions to first and second order
will then demand the equalities
\begin{equation}
\label{eq:pertmatch}
\hfamp^{(1)+}=\hfamp^{(1)-}, \quad \kfamp^{(1)+}=\kfamp^{(1)-},\quad
\hfamp^{(2)+}=\hfamp^{(2)-}, \quad \kfamp^{(2)+}=\kfamp^{(2)-}.
\end{equation}
defined on $(\Sigma_0,\hfamp)$ at first and second order respectively.

\begin{figure}[t]
  \hspace{-1cm}%
  \newcommand{\svgwidth}{9cm}
\begingroup%
  \makeatletter%
  \providecommand\color[2][]{%
    \errmessage{(Inkscape) Color is used for the text in Inkscape, but the package 'color.sty' is not loaded}%
    \renewcommand\color[2][]{}%
  }%
  \providecommand\transparent[1]{%
    \errmessage{(Inkscape) Transparency is used (non-zero) for the text in Inkscape, but the package 'transparent.sty' is not loaded}%
    \renewcommand\transparent[1]{}%
  }%
  \providecommand\rotatebox[2]{#2}%
  \ifx\svgwidth\undefined%
    \setlength{\unitlength}{655.20004883bp}%
    \ifx\svgscale\undefined%
      \relax%
    \else%
      \setlength{\unitlength}{\unitlength * \real{\svgscale}}%
    \fi%
  \else%
    \setlength{\unitlength}{\svgwidth}%
  \fi%
  \global\let\svgwidth\undefined%
  \global\let\svgscale\undefined%
  \makeatother%
  \begin{picture}(1,0.56043952)%
    \put(0,0){\includegraphics[width=\unitlength]{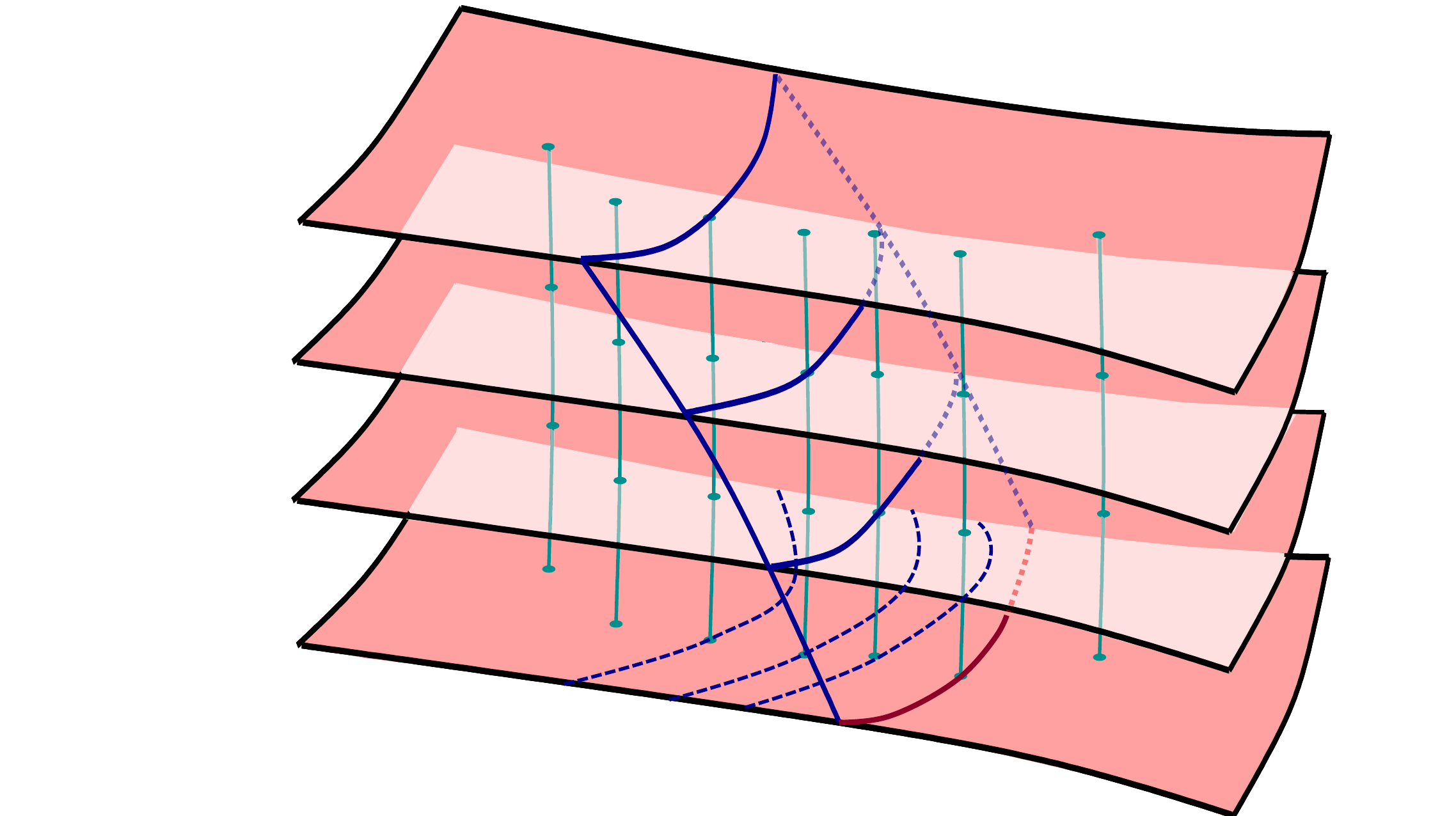}}%
    \put(0.92977028,0.1770475){\color[rgb]{0,0,0}\makebox(0,0)[lb]{\smash{$(\mmw_0,g)$}}}%
    \put(0.91444281,0.48209589){\color[rgb]{0,0,0}\makebox(0,0)[lb]{\smash{$(\mmw_\pertp,\gfam_\pertp)$}}}%
    \put(0.31954979,0.47218564){\color[rgb]{0,0,0}\makebox(0,0)[lb]{\smash{$\psi_\pertp$
}}}%
    \put(0.47564872,0.47216814){\color[rgb]{0,0,0}\makebox(0,0)[lb]{\smash{$\Supfam_\pertp$}}}%
    \put(0.70975703,0.15508383){\color[rgb]{0,0,0}\makebox(0,0)[lb]{\smash{$\Supfamp_0$}}}%
    \put(0.359383,0.04481528){\color[rgb]{0,0,0}\makebox(0,0)[lb]{\smash{$\Supfamp_{\pertp}$}}}%
  \end{picture}%
\endgroup%
  \newcommand{\svgwidth}{9cm}%
  \hspace{-1cm}%
\begingroup%
  \makeatletter%
  \providecommand\color[2][]{%
    \errmessage{(Inkscape) Color is used for the text in Inkscape, but the package 'color.sty' is not loaded}%
    \renewcommand\color[2][]{}%
  }%
  \providecommand\transparent[1]{%
    \errmessage{(Inkscape) Transparency is used (non-zero) for the text in Inkscape, but the package 'transparent.sty' is not loaded}%
    \renewcommand\transparent[1]{}%
  }%
  \providecommand\rotatebox[2]{#2}%
  \ifx\svgwidth\undefined%
    \setlength{\unitlength}{655.20004883bp}%
    \ifx\svgscale\undefined%
      \relax%
    \else%
      \setlength{\unitlength}{\unitlength * \real{\svgscale}}%
    \fi%
  \else%
    \setlength{\unitlength}{\svgwidth}%
  \fi%
  \global\let\svgwidth\undefined%
  \global\let\svgscale\undefined%
  \makeatother%
  \begin{picture}(1,0.56043952)%
    \put(0,0){\includegraphics[width=\unitlength]{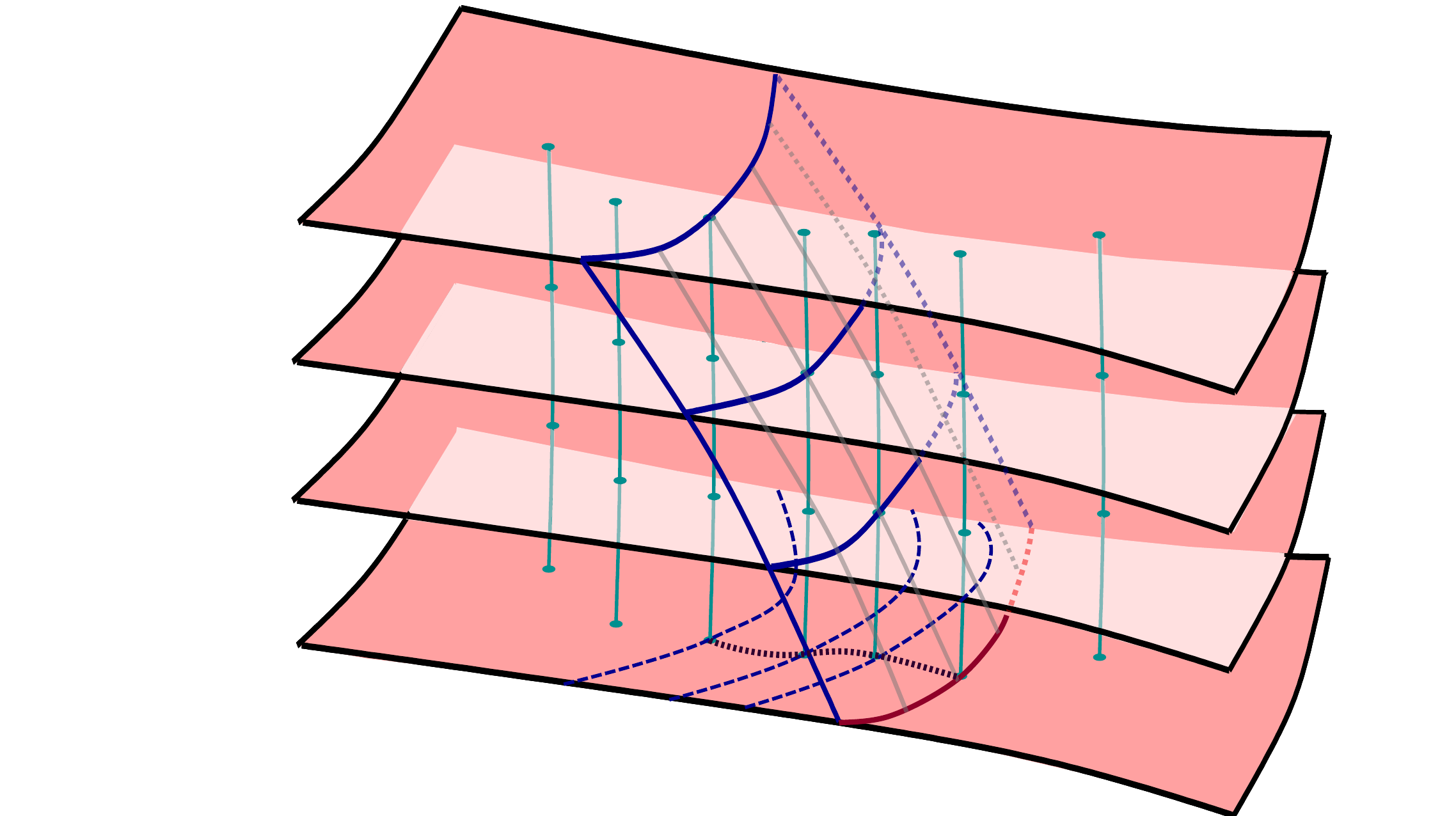}}%
    \put(0.31954979,0.47218564){\color[rgb]{0,0,0}\makebox(0,0)[lb]{\smash{$\psi_\pertp$
}}}%
    \put(0.47564872,0.47216814){\color[rgb]{0,0,0}\makebox(0,0)[lb]{\smash{$\Supfam_\pertp$}}}%
    \put(0.70975703,0.15508383){\color[rgb]{0,0,0}\makebox(0,0)[lb]{\smash{$\Supfamp_0$}}}%
    \put(0.6785623,0.08188377){\color[rgb]{0,0,0}\makebox(0,0)[lb]{\smash{$p$}}}%
    \put(0.359383,0.04481528){\color[rgb]{0,0,0}\makebox(0,0)[lb]{\smash{$\Supfamp_\pertp$}}}%
    \put(0.45310007,0.43338301){\color[rgb]{0,0,0}\makebox(0,0)[lb]{\smash{$\phi_\pertp$}}}%
    \put(0.56171047,0.13001889){\color[rgb]{0,0,0}\makebox(0,0)[lb]{\smash{$\gamma_p(\pertp)$}}}%
    \put(0.4455791,0.13217698){\color[rgb]{0,0,0}\makebox(0,0)[lb]{\smash{$p_\pertp$}}}%
  \end{picture}%
\endgroup%
  \caption{Family of spacetimes $(\mmw_\pertp,\gfam_\pertp)$, identified through the spacetime gauge
$\psi_\pertp$, with embedded hypersurfaces $\Supfam_\pertp$, identified, in turn, through the
hypersurface gauge $\phi_\pertp$. The projections of $\Supfam_\pertp$ onto the background
$(\mmw_0,g)$ via $\psi_\pertp$ are $\Supfamp_\pertp$. Given $p\in\Supfamp_0$ the composition
$\psi^{-1}_\pertp\circ\phi_\pertp(p)$ defines the path $\gamma_p(\pertp)$ on $\mmw_0$.
} 
\label{pic:pert_matching}
\end{figure}

The setting for the
construction of the tensors $\hfamp^\pm_\pertp$ and $\kfamp^\pm_\pertp$
on $(\Sigma_0,\hfamp)$ is described as follows. 
Take one side, say $+$, and assume each 
$\mmm^+_\pertp$ is a submanifold with boundary $\Supfam^+_\pertp$
in a larger $\mmw^+_\pertp$ with no boundary.
Hence, for each $\pertp$, $\Supfam_\pertp^+$ is an embedded hypersurface on $\mmw^+_\pertp$.
Identify the family $\mmw^+_\pertp$ pointwise
through $\psi^+_\pertp$ (extended appropriately)
to a background $\mmw^+_0\defi\mmw^+_{\pertp=0}$.
Each $\Supfam^+_\pertp$ is projected onto $\mmw^+_0$ via $\psi^+_\pertp$
in order to define an $\pertp$-family of hypersurfaces $\{\Supfamp^+_\pertp\}$
on $\mmw^+_0$, see the left side of Figure \ref{pic:pert_matching}.
This family describes how the background $\Supfamp^+_0$ changes
as $\pertp$ varies as a set of points on $\mmw_0^+$
with respect to the gauge $\psi_\pertp^+$.
But this is not enough to take $\pertp$-derivatives.
We still need to prescribe how a given
point $p\in\Sigma^+_0$ is mapped onto $\Supfamp^+_\pertp$.
For that we need to prescribe first an identification $\phi_\pertp:\Supfam_0\to\Supfam_\pertp$
for the family $\{\Supfam_\pertp\}$.
That comprises an additional gauge freedom, the so-called
\emph{hypersurface gauge freedom} \cite{Mukohyama00,Mars2005}.
The diffeomorphism $\phi_\pertp$ infers trivially another $\phi^+_\pertp$
for the family $\{\Supfam^+_\pertp\}$ through the embeddings on their respective $\mmw^+_\pertp$.
The composition of $\phi^+_\pertp$, from $p\in\Sigma^+_0$ to
$\Supfam^+_\pertp$, and $\psi^+_\pertp$ (down to $\Supfamp^+_\pertp$)
defines a path $\gamma_p(\pertp)$ in $\mmw_0$ starting at $p$
(see Figure \ref{pic:pert_matching}). The tangent vector to that path
at any $p\in\Supfamp^+_0$ and its acceleration 
define two vector fields $\vec Z^+_1$ and $\vec Z^+_2$, respectively, on $\Supfamp^+_0$.
The subscripts 1 and 2 refer to the fact that $\vec Z^+_1$ carries the information of the
deformation of $\Supfamp^+_0$ at first order, and $\vec Z^+_2$ at second order.
These are the so-called perturbation vectors of $\Sigma^+_0$ \cite{Mars2005}
(see also \cite{Battye01,Mukohyama00} for the first order).
The vectors $\vec Z^+$ (we refer to both $\vec Z^+_1$ and $\vec Z^+_2$)
depend on both
the spacetime and the hypersurface gauges by construction.
Let now $\vec n^+$ be a unit normal vector to $\Sigma^+_0$.
Every $\vec Z^+$ can thus be decomposed into 
normal and tangent parts, i.e.
\begin{equation}
\vec{Z}^+=Q^+ \vec n^+ + \vec T^+,
\label{eq:Zetas}
\end{equation}
where $\vec T^+$ is tangent to $\Sigma^+_0$.
The information on how the hypersurfaces $\Supfamp^+_\pertp$
vary as sets of points in $\mmm_0^+$
is carried only by $Q^+$, while $\vec T^+$ indicates
how the different points within those sets are identified.

The full calculation of the tensors $\hfamp^{(1)}$, $\kfamp^{(1)}$, $\hfamp^{(2)}$ and $\kfamp^{(2)}$
(let us drop the $+$ subscripts here)
in terms of the background configuration quantities
plus $\fpt$, $Q_1$, $\vec T_1$, and 
$\spt$, $Q_2$, $\vec T_2$
was performed in \cite{Mars2005} (Propositions 2 and 3),
and previously in \cite{Battye01,Mukohyama00} up to first order.
We include their expressions in Appendix \ref{sec:proofs}.

The picture discussed above makes apparent that
$\vec T_1$ and $\vec T_2$ fully depend on the hypersurface gauge
(as well as the spacetime gauges at either side).
It is important to stress that since $Q_1$ and $Q_2$ depend on the
spacetime gauge, the ``deformation'' they describe 
must be understood 
with respect to the spacetime gauge being used.
We will make use of the explicit transformations
of the perturbation vectors 
$\vec{Z}_1$ and $\vec{Z}_2$ under
spacetime gauges defined by $\vec s_1$ and $\vec s_2$.
These were shown in \cite{Mars2005} to be
\begin{equation}
\vec{Z}^{g}_1=\vec{Z}_1-\vec s_1,\qquad
\vec{Z}^{g}_2=\vec{Z}_2-\vec s_2-2\nabla_{\vec{Z}_1}\vec s_1+2\nabla_{\vec s_1}\vec s_1.
\label{eq:Z_gauges}
\end{equation}

The perturbed matching conditions are shown to be
(\ref{eq:pertmatch}) in terms of the background configuration quantities and
$\fpt^\pm$, $Q_1^\pm$, $\vec T_1^\pm$, and  $\spt^\pm$, $Q_2^\pm$, $\vec T_2^\pm$
in Theorem 1 in \cite{Mars2005}.
It must be stressed that the tensors $h^{(1)\pm}$, $h^{(2)\pm}$
$\kfamp^{(1)\pm}$, $\kfamp^{(2)\pm}$ are spacetime
gauge invariant by construction, and thus conditions (\ref{eq:pertmatch}).
Moreover, although the tensors are not hypersurface gauge invariant,
the matching conditions (\ref{eq:pertmatch}) are, provided the background is matched
\cite{Mukohyama00,Mars2005}.
Let us emphasize that $Q^{\pm}_{1/2}$ and $\vec T^{\pm}_{1/2}$ are a priori
unknown quantities and fulfilling the matching conditions
requires {\it showing} that two pairs of vectors $\vec Z^\pm_{1/2}$ exist such that
(\ref{eq:pertmatch}) are satisfied. The spacetime gauge freedom at either side can be exploited to 
fix either or both pairs  $\vec Z^{+}_{1/2}$ or $\vec Z^{-}_{1/2}$ independently a priori,
but this has to be carefully analyzed
if additional spacetime gauge choices are made.
Finally, the hypersurface gauge is common to both sides, and therefore, it can be used to
fix one of the vectors $\vec T^{+}$ or $\vec T^{-}$, but not both
(at first and second order).

At either side, say $+$, we will call a gauge $\psi^+_\pertp$ ``surface-comoving'' if
the hypersurfaces $\Supfamp^+_\pertp$ do not vary, and thus agree with $\Supfamp_0^+$,
as sets of points in $\mmm_0^+$.
At first order that is equivalent to $Q_1^+=0$, but at second order
$Q_2^+$ carries more information coming from the first order.
This fact will motivate
the introduction of the quantity $\Qt2$ in Section \ref{sec:second_order_matching}.
The gauges referred to as ``surface gauges'' in previous
works, e.g. \cite{Brizuela2010,JMMG-Gundlach},
require the vanishing of the whole perturbation vector $\vec Z$.

\subsection{Family of metrics}
Although the original ``perturbed'' metric in \cite{Hartle1967} is given by
(\ref{metricHartle}) assuming also that $\kH$ has no $l=0$ term, i.e. in the (spacetime) $k$-gauge,
the determination
of the matching hypersurface is made  in \cite{Hartle1967}
(and most other works in the literature)
by resorting to another spacetime gauge,
prescribed through the surfaces of constant energy density. Since we also want
to examine the use of these different spacetime gauges in the literature we consider
a family of metrics $\{\gfamp_\pertp\}$ that can accommodate both spacetime gauges.
To do that a crossed term in $(r,\theta)$ is needed.

Let us thus define the following one-parameter family 
$\{\gfamp_\pertp\}$ on $(\mathcal{V}_0,\gback)$, where $\gback=\gfamp_{\pertp=0}$,
taken up to order $\pertp^2$
\begin{eqnarray}
\gfamp_{\pertp} &=& -e^{\nu(r)}\left(1+2 \pertp^2 h(r,\theta)\right)dt^2 + e^{\lambda(r)}\left(1+ 2 \pertp^2 m(r,\theta) \right)dr^2 +2r e^{\lambda(r)}\pertp^2 \partial_\theta f(r,\theta) dr d\theta\nonumber \\
&& + r^2(1+2 \pertp^2 k(r,\theta))\left[d\theta^2 + \sin ^2 \theta (d\varphi-\pertp \omega(r,\theta) dt)^2\right]\,
+ \mathcal{O}(\pertp^3), \label{gefamily}
\end{eqnarray}
where $t\in(-\infty,\infty)$, $r>0$, $\theta\in(0,\pi)$ and $\varphi\in[0,2\pi)$.
Clearly, an arbitrary function of $r$ 
can be added to $f(r,\theta)$ with no consequences.
The appearance of $\fcross$ differentiated is just a mere convenience.
$\{\gfamp_{\pertp}\}$ is a family of stationary and axisymmetric metrics on $(\mathcal{V}_0,\gback)$. The (unique) axial Killing vector field will be denoted by $\vec \axial=\partial_\varphi$, and
we will single out the timelike Killing $\vec \stat=\partial_t$.  
The first and second order metric perturbation tensors, $\fpt= \left.\partial_\pertp g_\pertp\right|_{\pertp=0}$ and $\spt =\left. \partial_\pertp^2 g_\pertp\right|_{\pertp=0}$ respectively, take thus the form
\begin{eqnarray}
\fpt  &=& 
 -2r^2\, \omega(r,\theta) \sin ^2 \theta dt d\varphi \label{fopert_tensor},\\
\spt &=& \left(-4 e^{\nu(r)} h(r, \theta) + 2r^2 \sin ^2 \theta {\omega}^2(r, \theta)\right)dt^2 + 4 e^{\lambda(r)} m(r, \theta) dr^2 \nonumber\\
&&+4 r^2 k(r, \theta)
(d\theta^2 + \sin ^2 \theta d\varphi^2)
+ 4re^{\lambda(r)}\partial_\theta f(r,\theta)dr d\theta ,\label{sopert_tensor}
\end{eqnarray}
defined on the spherically symmetric and static spacetime
background $(\mathcal{V}_0,\gback)$ with
\begin{equation}
\gback= -e^{\nu(r)} d{t}^ 2 + e^{\lambda(r)} d{r}^ 2+{r}^ 2(d{\theta}^2+\sin^2 \theta d {\varphi}^ 2).\label{eq:g0}
\end{equation}

The (spacetime) gauge transformations described by $\vec s_1=Ct\partial_\varphi$, with arbitrary
constant $C$, at first order and $\vec V_2=2S(r,\theta)\partial_r$, for an arbitrary $S(r,\theta)$,
are contained within the family $g_\pertp$. Under the gauge $\vec s_1=Ct\partial_\varphi$, the perturbation tensor $\fpt$
transforms as (\ref{eq:Ksgauged})
\begin{equation}
\fpt^{g}=-2r^2\, (\omega-C) \sin ^2 \theta dt d\varphi,
\label{eq:K1_s1}
\end{equation}
while under a change $\vec V_2=2S(r,\theta)\partial_r$ (with $\vec s_1=Ct\partial_\varphi$),
$\spt$ transforms as (\ref{eq:Ksgauged})
\begin{eqnarray}
\spt^{g} 
&=&\left(-4e^{\nu}\left(h + \frac{\nu'}{2}S\right)+2r^2\sin^ 2 \theta (\omega-C)^2 \right)dt^2+4e^{\lambda}\left(m +e^{-\frac{\lambda}{2}}\left(S e ^{\lambda/2} \right)'\right)dr^2\nonumber \\
&&+4r^2\left( k +\frac{S}{r}\right)(d\theta^2 + \sin ^2 \theta d\varphi^2) +4r e^ {\lambda}\partial_\theta \left(f+\frac{S}{r} \right)dr d\theta.\label{eq:K2_s2}
\end{eqnarray}
We will refer to this class of second order
gauge transformations as `radial' gauges. 

A (spacetime) gauge whitin the set of these 'radial' gauges will be fixed, partially or completely,
whenever the functions appearing in $\fpt$, (\ref{fopert_tensor}), and/or $\spt$, (\ref{sopert_tensor}),
are restricted in any way. The remaining freedom would consist on the possible $C$ and $S(r,\theta)$
that make the changes to the components of (\ref{eq:K1_s1}) and (\ref{eq:K2_s2}) fit, component-wise, within that restriction.
The $k$-gauge, as mentioned, consists of imposing that the function $k(r,\theta)$ in (\ref{sopert_tensor})
has no $l=0$ part, and that $f=0$. In that case, the restriction on the $\spt_{\theta\theta}$ component
implies that $S(r,\theta)$ cannot have $l=0$ part, while the restriction on the $\spt_{r\theta}$ component
needs that $S(r,\theta)$ does not depend on $\theta$. The only possibility is thus $S(r,\theta)=0$, so that there is
no freedom left. We thus say that the $k$-gauge fixes completely the 'radial' gauge.

Let us now consider a couple of
background spacetimes $(\mathcal{V}^\pm_0,\gback^\pm)$, with corresponding coordinates
$\{t_\pm,r_\pm,\theta_\pm,\varphi_\pm\}$
and families of metrics $g_\pertp^\pm$ as given in (\ref{gefamily}). 
Greek indices will denote quantities defined on $\mathcal{V}^\pm_0$.
In what follows we present 
the perturbed matching over a spherically symmetric (and static) background configuration
composed by the matching of $(\mathcal{V}^+_0,\gback^+)$ and $(\mathcal{V}^-_0,\gback^-)$.
Let us note that by using (\ref{gefamily}) (at both sides)
we will be implicitly assuming that the perturbation will be performed within the family of (spacetime) gauges for which
(\ref{fopert_tensor})-(\ref{sopert_tensor}) hold.
We will not be using the field equations until Section \ref{sec:pf-vacuum}.

The structure of the original metric (\ref{metricHartle}) can
be clearly recovered by taking $\fcross=0$
and noting that the choice of perturbation parameter $\pertp$
is not relevant, since families of solutions are obtained by scaling.
The physics of the model will restrict the scalability (see Eq. (1) in \cite{Hartle1967}).
Note, however, that the relation between the radial coordinates in (\ref{metricHartle})
and (\ref{gefamily}) (either $r_\pm$) must still be determined in order to be able to compare
the functions in (\ref{metricHartle}) and (\ref{gefamily}).
That is the purpose of the concluding Section \ref{sec:conclusion}.

\subsection{Background configuration}
The background configuration is chosen to be globally spherically symmetric and  static.
This translates to the fact that the matching of $(\mathcal{V}^+_0,\gback^+)$ and $(\mathcal{V}^-_0,\gback^-)$,
through respective boundaries $\Sigma_0^+$ and $\Sigma_0^-$, is asked to 
preserve the symmetries (see \cite{Vera2002}), both the spherical symmetry and staticity. Under that condition the
hypersurfaces $\Sigma_0^+$ and $\Sigma_0^-$
to be matched can be finally cast as (see e.g. \cite{Mars2007})
\begin{eqnarray}
\Sigma_0^+ &=& \{\tp= \tau, \rrp=\ro , \thetap = \vartheta, \phip = \phi\},\label{sigma0+}\\
\Sigma_0^- &=& \{\tm= \tau, \rrm=\ro , \thetam = \vartheta, \phim = \phi\},\label{sigma0-}
\end{eqnarray}
for a constant $\ro>0$, without loss of generality. The coordinates
$\{\tau,\vartheta,\phi\}$ parametrize the abstract manifold
$\Sigma_0\equiv \Sigma_0^+= \Sigma_0^-$.
Latin indices $i,j,\ldots$ will refer to objects on $\Sigma_0$.

The tangent vectors to $\Sigma_0^+$ and $\Sigma_0^-$ thus read
\begin{equation}
\vec e^\pm_i:\qquad \vec e^\pm_1=\partial_{t_\pm}|_{\Sigma^\pm_0},\quad \vec e^\pm_3=\partial_{\theta_\pm}|_{\Sigma^\pm_0},\quad \vec e^\pm_2=\partial_{\varphi_\pm}|_{\Sigma^\pm_0}, 
\label{eq:tangents}
\end{equation}
and the corresponding unit normals 
are 
\begin{equation}
\vec{n}^+= -e^{-\frac{\lambda_+(\ro)}{2}}\partial_{\rrp}|_{\Sigma^+_0},\quad
\vec{n}^-= -e^{-\frac{\lambda_-(\ro)}{2}}\partial_{\rrm}|_{\Sigma^-_0},\label{normal_vector}
\end{equation}
under the condition that $\vec n^+$ points $\mathcal{V}_0^+$ inwards
and $\vec n^-$ points $\mathcal{V}_0^-$ outwards, so that as $\rrp$
increases one reaches $\mathcal{V}_0^-$, and as $\rrm$ increases one
gets away of $\mathcal{V}_0^+$.  This convention will be used in what
follows in order to call $\mathcal{V}_0^+$ the \emph{interior} and
$\mathcal{V}_0^-$ the \emph{exterior}.
Clearly $\Supfamp_0^\pm$ are timelike hypersurfaces everywhere, and are
(equally) oriented by construction.

The first and second fundamental forms read
\begin{eqnarray*}
\hfamp^\pm_{ij}dx^idx^j=-e^{\nu_\pm(\ro)}d\tau^2 +\ro^2(d\vartheta^2+\sin^2\vartheta d\phi^2),\label{h0ij}\\
\kfamp^\pm_{ij}dx^idx^j=e^{-\frac{\lambda_\pm(\ro)}{2}}\left(\frac{1}{2}e^{\nu_\pm(\ro)}\nu'_\pm(\ro)d\tau^2-\ro(d\vartheta^2+\sin^2\vartheta d\phi^2)\right),
\end{eqnarray*}
where a prime denotes differentiation with
respect to the corresponding radial coordinate, i.e. $\rrp$ or $\rrm$ accordingly.
The matching conditions $\hfamp^+=\hfamp^-$ and $\kfamp^+=\kfamp^-$  are thus equivalent to
\begin{equation}
[\nu]=0,\quad [\nu']=0,\quad [\lambda]=0, \label{background_matching}
\end{equation}
where we follow the usual notation
$[f]=f^+|_{\Sigma^+_0}-f^-|_{\Sigma^-_0}$ for objects $f^\pm$
defined at either side.
For the sake of brevity, given a pair $f^\pm$ satisfying $[f]=0$, we will
simply denote by $f|_{\Sigma_0}$ either of the equivalent $f^+|_{\Sigma^+_0}$
or $f^-|_{\Sigma^+_0}$.
The background matching hypersurface $\Supfamp_0$ is endowed with the metric
$h=-e^{\nu(\ro)}d\tau^2 +\ro^2(d\vartheta^2+\sin^2\vartheta d\phi^2)$.

Once the static and spherically symmetric background configuration has been constructed
we proceed to study the perturbed matching up to second order.
As discussed above,
the ingredients needed are the tensors that describe the perturbations at either side, i.e.
the first and second metric perturbation tensors $\fpt^\pm$ and $\spt^\pm$ as defined above
(\ref{fopert_tensor})-(\ref{sopert_tensor}),
plus the two
(so far unknown) perturbation
vectors $\vec{Z}_1^\pm$ and $\vec{Z}_2^\pm$
given in the form (\ref{eq:Zetas}). 
To ease the notation we will denote by $Q^\pm$ and $\vec
T^\pm=T^\tau_\pm(\tau,\vartheta,\phi)\partial_\tau+T^\vartheta_\pm(\tau,\vartheta,\phi)\partial_\vartheta+T^\phi_\pm (\tau,\vartheta,\phi)\partial_\phi$
both the objects defined on each $\mmm^\pm_0$ and
the corresponding pullback and pushforward quantities that live on
$\Sigma_0$.  The same applies for the
functions in (\ref{fopert_tensor})-(\ref{sopert_tensor}), which will
be denoted equivalently as functions restricted to points on $\Sigma^\pm_0 \subset
\mathcal{V}^\pm_0$ and functions on $\Sigma_0$ whenever that does not lead to
confusion.
It is not difficult to show that
the fact that since the final perturbed matched spacetime is assumed to preserve the axial symmetry the functions $Q$ and components
$T^i$ do not depend on $\phi$.
Nevertheless, we will take that as an assumption.
The first and second order perturbed matchings are ruled by the particularisation of
Theorem 1 together with Propositions 2 and 3 from \cite{Mars2005}
to the present setting with the above ingredients.
For completeness, the explicit expressions of the first and second order first and second
fundamental forms are included in Appendix \ref{sec:proofs}.

\subsection{First order matching}
\label{sec:first_order_theorem}

\begin{proposition}\label{teo:first_order_matching}
Let $(\mathcal{V}_0,\gback)$ be the static and spherically symmetric
spacetime resulting from the matching of $(\mathcal{V}_0^+,\gback^+)$
and $(\mathcal{V}_0^-,\gback^-)$, with $g^\pm$ given by (\ref{eq:g0}) with respective $\pm$ in functions and coordinates,
across $\Sigma^\pm_0$, defined by
(\ref{sigma0+}), (\ref{sigma0-}), with $\ro>0$, so that the matching conditions
(\ref{background_matching}) hold and the unit normals
(\ref{normal_vector}) are chosen following the above interior/exterior
convention. 
Consider the metric perturbation tensors $\fpt^\pm$ as defined in
(\ref{fopert_tensor}) at either side $\mathcal{V}^\pm_0$, plus two unknown
functions $Q_1^\pm(\tau,\vartheta)$ and two unknown vectors
$\vec{T}^\pm_1=T_1^\pm{}^\tau(\tau, \vartheta) \partial_\tau
+T_1^\pm{}^\vartheta (\tau, \vartheta) \partial_\vartheta+
T_1^\pm{}^\phi(\tau, \vartheta) \partial_\phi $ on $\Sigma_0$.

The necessary and sufficient conditions that  $\fpt^\pm$ must
satisfy to fulfil the first order matching conditions are 
\begin{eqnarray}
\left[\omega \right]&=&b_1,\label{omega_matching}\\
\left[\omega'\right]&=&0, \label{omegap_matching}
\end{eqnarray}
where $b_1$ is an arbitrary constant. Regarding the perturbed matching hypersurface,
if 
\begin{equation}
2e^{\lambda(\ro)}-2+\ro\nu'(\ro)\neq 0
\label{condition_f}
\end{equation}
the remaining first order matching conditions read 
\begin{eqnarray}
&&[\vec{T}_1]= b_1 \tau \partial_\phi ,\label{T1_matching}\\
&&\left[Q_1\right]=0,\qquad Q_1[\lambda']=0\qquad  Q_1[\nu'']=0. \label{Q1_matching} 
\end{eqnarray}
\end{proposition}

The proof is left to Appendix \ref{sec:proofs}.
Note that although $[Q_1]=0$ is always a necessary condition, $(Q_1^\pm=)Q_1=0$ is not. Indeed,
if the background configuration satisfies $[\lambda']=0$ and $[\nu'']=0$,
$Q_1$ can be any arbitrary function of $(\tau, \vartheta)$.
Finally, let us remark that condition (\ref{condition_f}) will be satisfied in all
cases we will be interested in.

\subsection{Second order matching}
\label{sec:second_order_matching}
Let us first define at each side $\Sigma_0^\pm$ the following 
quantity
\begin{eqnarray}
&&\Qt2 := \Qtwo+\kfamp_{\alpha \beta}T_1^\alpha T_1^\beta-2\vec{T}_1(Q_1)\nonumber\\
&&= \Qtwo + a e^{-\lambda(a)/2}\left\lbrace  \frac{\nu'e^\nu}{2a}{(T_1^\tau)}^2 - \sin^2 \vartheta {(T_1^\phi)}^2 - {(T_1^\vartheta)}^2 \right \rbrace - 2 (T_1^\tau \partial_\tau Q_1 + T_1^\vartheta \partial_\vartheta Q_1), 
\label{eq:def_Qt2}
\end{eqnarray} 
which, given the above first order matching conditions (\ref{T1_matching}), leads to
\[
[ \Qt2 ]= [ \Qtwo ] + a e^{-\lambda(a)/2}  \sin^2 \vartheta b_1 \tau \left(b_1 \tau - 2 T_1^+{}^\phi \right)
- 2 (T_1^\tau \partial_\tau [Q_1] + T_1^\vartheta \partial_\vartheta [Q_1]).
\]
This new $\Qt2$ will substitute the original $\Qtwo$ 
in this section.
The most immediate purpose for introducing this quantity is to absorb some first order terms arising
from the matching equations, 
and thus keep more compact expressions.\footnote{It is not difficult to check that $\Qt2$ is \emph{hypersurface} gauge invariant.}

\begin{proposition} \label{teo:second_order_matching}
Let $(\mathcal{V}_0,\gback)$ with $\Sigma_0$ be the static and
spherically symmetric background matched spacetime as described in
Proposition \ref{teo:first_order_matching}, and assume that
(\ref{condition_f}) is satisfied.  Let it be perturbed to first order
by $\fpt^\pm$ plus $Q_1^\pm$ and $\vec T_1^\pm$ so that
(\ref{omega_matching}), (\ref{omegap_matching}), (\ref{T1_matching}),
(\ref{Q1_matching}) hold.  Consider the second order metric
perturbation tensor $\spt^\pm$ as defined in (\ref{sopert_tensor}) at
either side, plus two unknown functions $\Qt2^\pm(\tau,\vartheta)$ and
two unknown vectors
$\vec{T}^\pm_2=T_2^\pm{}^\tau (\tau,\vartheta) \partial_\tau 
+T_2^\pm{}^\vartheta (\tau,\vartheta) \partial_\vartheta
+ T_2^\pm{}^\phi (\tau,\vartheta) \partial_\phi$ on $\Sigma_0$.

If either $\left[\lambda'\right]\neq 0$ or $\left[\nu''\right]\neq 0$, so that $(Q^\pm_1=)Q_1=0$,
the necessary and sufficient conditions that $\spt^\pm$ must
satisfy to fulfil the second order matching conditions  
are
\begin{eqnarray}
  &&\difk=c_1 \cos\vartheta+c_2+\difg\label{eq:mc:k} \\
  &&\difh=\frac{1}{2}H_0+\frac{1}{4}\ro \nu'(a)\left\{2\difk+H_1\cos\vartheta\right\}\label{eq:mc:h}\\
  &&\difm=\ro \difkp+\frac{1}{4}e^{-\lambda(\ro)/2}\left[\lambda'\right]\Qt2^+
  +\frac{1}{4}\left(\ro\lambda'_-(\ro)+2\right)\left\{2\difk + H_1\cos\vartheta\right\}\nonumber\\
  &&\qquad-\frac{1}{2}(H_1+2c_1)e^{\lambda(\ro)}\cos\vartheta \label{eq:mc:m}\\
  &&\difhp=\frac{1}{2}\ro \nu'(\ro) \difkp+\frac{1}{4}e^{-\lambda(\ro)/2}\left[\nu''\right]\Qt2^+
  +\frac{1}{4}\left(\ro\nu''_-(\ro)+\nu'(\ro)\right)\left\{2\difk + H_1\cos\vartheta\right\}\nonumber\\
  &&\qquad-\frac{1}{4}(H_1+2c_1)\nu'(\ro)e^{\lambda(\ro)}\cos\vartheta,\label{eq:mc:hp}
\end{eqnarray}
for arbitrary constants $c_1$, $c_2$, $H_0$ and $H_1$ and function $\Qt2^+(\vartheta)$.

If $\left[\lambda'\right]=0$ and $\left[\nu''\right]= 0$, then $[\omega'']Q_1=0$ and the above equations are the same except for two changes
in (\ref{eq:mc:m}) and (\ref{eq:mc:hp}) given respectively by
\begin{equation}
\left[\lambda'\right]\Qt2^+\to -e^{-\lambda/2}\left[\lambda''\right] (Q_1)^2,
\qquad
\left[\nu''\right]\Qt2^+\to -e^{-\lambda/2}\left[\nu'''\right] (Q_1)^2.
\label{eq:changes_for_Q1}
\end{equation}
In all cases, the relation 
\begin{equation}
  \label{eq:mc:Q2}
  \left[\Qt2\right]=\ro e^{\lambda(\ro)/2}\left\{2\difk+ H_1\cos\vartheta\right\}
\end{equation}
must hold, hence $\left[\Qt2\right]$ cannot depend on $\tau$,
and the differences $[\vec T_2^\pm]$ satisfy 
\begin{eqnarray}
\left[T_2^\tau\right]&=&-H_0\tau,\nonumber\\
\left[T_2^\phi\right]&=& 2b_1(T_1^\tau + \tau T_1^\vartheta \cot\vartheta)
-\frac{2}{\ro}e^{-\lambda(\ro)/2}b_1\tau Q_1^+  ,\label{eq:difz2p}\\
\left[T_2^\vartheta\right] &=& \left(b_1 \tau \cos \vartheta(b_1\tau - 2T_1^+{}^\phi)+H_1\right)\sin \vartheta.\nonumber
\end{eqnarray}

\end{proposition}

The proof is left to Appendix \ref{sec:proofs}.
Let us remark that in the $Q_1\neq 0$ case, the corresponding equations for $\difm$ and
$\difh$, (\ref{eq:mc:m}) and (\ref{eq:mc:hp}) with the corresponding
changes (\ref{eq:changes_for_Q1})
(see (\ref{app:difm}) and (\ref{app:difhp}) in Appendix \ref{sec:proofs})
imply that if $[\lambda'']\neq 0$ or $[\nu''']\neq 0$ then $Q_1$
cannot depend on $\tau$.
On the other hand, the condition $[\omega'']Q_1=0$ will be automatically satisfied
in all cases of interest, once the field equations are imposed, as shown below.

\section{Hartle's setting for a perfect fluid interior and vacuum exterior}
\label{sec:pf-vacuum}
In this section we focus on a global configuration composed of a
rigidly rotating perfect fluid ball (with no convective motions)
immersed in an asymptotically flat vacuum
exterior. To present the equations in this section
we will drop the $+$ and $-$ symbols in most places if they are not necessary.
Both regions can be considered
to be of perfect fluid type, from which the vacuum case is recovered trivially.

Let us then impose the metrics $\gfam_{\pertp}$ to satisfy
the equations
$
\hat G(\hat g_\pertp)_{\alpha\beta}=8\pi \hat T_\pertp{}_{\alpha\beta}
$
for an energy momentum tensor of the form 
\[
  \hat T_{\pertp} = ( \hat \energy_\pertp + \hat \pressure_\pertp) \hat u_\pertp \otimes\hat u_\pertp + \hat\pressure_\pertp \gfam_{\pertp{}},
\]
where $\hat u_\pertp$ is the (unit) fluid flow,
and $\hat \energy_\pertp$ and $\hat \pressure_\pertp$, eigenvalues of $\hat T_\pertp$, the corresponding
mass-energy density and pressure. 
Note that the fluid vector $\hat u_\pertp$ and corresponding ``hatted'' scalars are objects defined, still,
on each $(\mmm_\pertp,\gfam_\pertp)$. All these objects, in \textit{covariant form},
are now pulled back through $\psi_\pertp$ down onto $(\mmm_0,g)$ (see Section \ref{sec:overview}).
That defines the associated families of (tensorial) objects $G_\pertp{}$, $T_{\pertp}{}$,
$\energy_\pertp$, $\pressure_\pertp$ and $U_\pertp$ on $(\mmm_0,g)$, which therefore satisfy
\begin{equation}
 G( g_\pertp)_{\alpha\beta}=8\pi  T_\pertp{}_{\alpha\beta}
\label{eq:EFEs}
\end{equation}
with
\begin{equation}
T_{\pertp} = ( \energy_\pertp + \pressure_\pertp) U_\pertp \otimes U_\pertp + \pressure_\pertp \gfamp_{\pertp{}},
\label{em-family}
\end{equation}
by construction. It is worth mentioning that the (families of) objects do depend on the gauge
defined by $\psi_\pertp$, and thus also the right and left hand sides of (\ref{eq:EFEs}). However,
the equations (\ref{eq:EFEs}) themselves do not depend on the gauge, in the sense that if
(\ref{eq:EFEs}) are fulfilled in one gauge, they will be satisfied in any other gauge.

On the other hand,
the fluid vector in contravariant form can also be pushforwarded through $\psi^{-1}_\pertp$,
and thus yet obtain another family of vectors $\vec u_\pertp$ on $(\mmm_0,g)$.
Since $\hat u_\pertp{}_\alpha \hat u_\pertp{}^\alpha=-1$
at each $(\mmm_\pertp,\gfam_\pertp)$,
we must have $U_\pertp{}_\alpha u_\pertp{}^\alpha=-1$ on $(\mmm_0,g)$.
This can be shown to be equivalent to $\gfamp_\pertp{}_{\alpha\beta} u_\pertp{}^\alpha u_\pertp{}^\beta=-1$,
and corresponds to the normalisation condition that  $\vec u_\pertp$ must satisfy.
We can take now $\pertp$-derivatives and construct the expansion of $\vec u_\pertp$ as
$\vec u_\pertp=\vec u+\pertp \vec u^{(1)}+ \frac{1}{2}\pertp^2 \vec u^{(2)}+ \mathcal{O}(\pertp^3)$,
and 
\begin{eqnarray}
\energy_\pertp &=& \Eb + \pertp \Ep + \frac{1}{2}\pertp^2 \Epp + \mathcal{O}(\pertp^3),\label{Energy_epsilon}\\
\pressure_\pertp &=& \Pb + \pertp \Pp + \frac{1}{2}\pertp^2 \Ppp+  \mathcal{O}(\pertp^3).\label{Pressure_epsilon}
\end{eqnarray}
All functions and vector components depend on $r$ and $\theta$.
We will consider later the existence of a barotropic equation of state for the $\pertp$-family,
independent of $\pertp$, so that
$P_\pertp$ is a function of $E_\pertp$ alone.
Taking $\pertp$-derivatives, such relation yields a constraint for
the first and second order expansions, which must satisfy, respectively
\begin{eqnarray}
\Pp - \frac{\partial P }{\partial E} \Ep = 0,\nonumber\\
\Ppp - \frac{\partial P }{\partial E} \Epp - \frac{\partial^2 P}{\partial E^2}{\Ep}^2  = 0\label{eq:2ndbarotropic}.
\end{eqnarray}

The absence of convective motions translates onto the condition that $\vec u_\pertp$
lies on the orbits of the group generated by $\{\vec\axial,\vec\stat\}$,
this is $\vec u_\pertp\propto\vec\stat+\kfamp(\pertp,r,\theta)\vec\axial$
for some function $\kfamp$.
Rigid rotation demands that $\kfamp(\pertp,r,\theta)$ does not depend on $\{r,\theta\}$, so that
$\vec u_\pertp$ are proportional to (timelike) Killing vector fields \cite{Exact_solutions}, i.e.
$\vec u_\pertp=N(\pertp)(\vec\stat+\kfamp(\pertp)\vec\axial)$
for some function $\kfamp(\pertp)$, with $N(\pertp)$ fixed by the above normalisation.
A static background configuration forces $\kfamp(0)=0$, and therefore
$\kfamp(\pertp)=\pertp\Omega+O(\pertp^2)$ for some constant $\Omega$. This constant $\Omega$
is gauge dependent (see below, in Section \ref{sec:gauges_first_order}).
Following \cite{Hartle1967} we assume that $\pertp$ drives a rotational peturbation, so that
only odd powers enter $\kfamp(\pertp)$.
In components we thus demand
\begin{equation}
  \label{eq:conv_rigid}
  u_\pertp^\varphi=\pertp\Omega u_\pertp^t,\qquad u_\pertp^r=u_\pertp^\theta=0.
\end{equation}
This (gauge-dependent) constant $\Omega$ differs from the perturbation parameter
(which we denote by $\OH$) as defined in \cite{Hartle1967}. In the present scheme the perturbation parameter $\pertp$
has been defined abstractly, a priori independently of the rotation parameter $\Omega$.
The translation will be given by $\OH=\pertp(\Omega+B)$,
where $B$ is a constant to be determined later, see Section \ref{sec:gauges_first_order}.

The vacuum equations are obtained by simply setting $\energy_\pertp=\pressure_\pertp=0$.
\subsection{Background}
The matter content of the interior region of the
background configuration is a perfect fluid,
static and spherically symmetric. Its normalized 4-velocity is $\vec
u=e^{-\nu/2}\partial_t$. The two field equations providing
$\Eb$ and $\Pb$ in terms of the metric functions are
\begin{eqnarray}
\lambda' &=& \frac{1}{r}(1-e^\lambda)+re^\lambda 8\pi\Eb,
\label{eq:lambdaprime}
\\
\nu' &=&\frac{1}{r}(e^\lambda-1)+re^\lambda 8\pi\Pb,\label{eq:nuprime}
\end{eqnarray}
while the pressure isotropy condition yields the equation
\begin{equation}
2 r \nu ''+\nu '(r \nu '-2)-\lambda'(2+r \nu ')+\frac{4}{r} \left(e^{\lambda }-1\right)=0, \label{eq:isotropy}
\end{equation}
which can be also written, using (\ref{eq:lambdaprime})-(\ref{eq:nuprime}), as
\begin{equation}
\Pb'=-\frac{1}{2r} (\Eb+\Pb)\left(e^\lambda-1+8\pi r^2 e^\lambda \Pb\right).\label{eq:Pprime}
\end{equation}

Let us now define $M(r)$ and $j(r)$, which will be useful for
the comparison of the expressions here with those in \cite{Hartle1967}, by 
\begin{eqnarray}
j(r) &\defi& e^{-(\lambda+\nu)/2},\nonumber \\
1-\frac{2M(r)}{r}&\defi& e^{-\lambda}.\label{eq:Mass}
\end{eqnarray}
In the vacuum region ($-$) the field equations (\ref{eq:isotropy}) imply that $M(\rrm)$ is a
constant, which will be denoted by $M$ as usual, and that 
\begin{equation}
e^{-\lambda_-(\rrm)} = e^{\nu_-(\rrm)} =1-\frac{2M}{\rrm} \qquad \Rightarrow j(\rrm)=1.\label{eqs_back_vacuum}\\
\end{equation}
We will assume $M>0$.
The matching of the background (\ref{background_matching})
implies, in particular, that
\begin{equation}
\nu_+(\ro)=-\lambda_+(\ro)=\log\left(1-\frac{2M}{\ro}\right),\label{eq:cont_back}
\end{equation}
and the following expressions for
the differences of the derivative of the functions of the metric
in terms of the fluid variables
\begin{eqnarray}
\left[\nu'\right] &=& \ro e^{\lambda(a)}8\pi[\Pb ]=0, \label{cont_pressure} \\
\left[\lambda'\right] & = & \ro e^{\lambda(\ro)} 8\pi [\Eb], \label{Bp} \\
\left[\nu''\right] &=& \left( 1+\frac{\ro\nu'(\ro)}{2}\right)\frac{[\lambda']}{\ro}=\left(1+\frac{\ro\nu'(\ro)}{2}\right)e^{\lambda(\ro)}8\pi[\Eb].\label{nupp}
\end{eqnarray}
Note that the difference $[\Eb]$
corresponds to the value of the interior energy density $\Eb_+$ on
$\Sigma_0$, this is $[\Eb]=\Eb_+(\ro)$, for a vacuum exterior. 
We just prefer to keep $[\Eb]$ in some expressions for the sake of generality, since they apply in the 
matching of two fluids, and the notation is, in fact, more compact.

It must be stressed that whereas the matching condition (\ref{cont_pressure}) implies,
for a vacuum exterior, that $\Pb (\rrp )$ must vanish on $\Supfamp_0$,
the energy density $\Eb(\ro)$ stays free, \emph{a priori}.
Its value will be determined, if any, by the equation of state.
Let us remark finally that the condition (\ref{condition_f})
of Proposition \ref{teo:first_order_matching} is now equivalent to $M\neq 0$.

\subsection{First order problem}
\label{sec:first_order}
The absence of convective motions and rigid
rotation (\ref{eq:conv_rigid}), together with the normalisation condition,
yield to first order
\begin{equation}
\vec{u}^{(1)} = \Omega u^t \partial_\varphi = \Omega e^{-\nu/2} \partial_\varphi. \label{rigidrotation}
\end{equation}
The linearized energy momentum tensor is found by
taking the $\pertp$-derivative of (\ref{em-family}) and evaluating on $\pertp=0$,
$
T^{(1)}_{\alpha \beta} = (\Ep+\Pp){u}_\alpha {u}_\beta
+\Pp \gback_{\alpha \beta}
+ 2(\Eb+\Pb)(u^{(1)}_{(\alpha}{u}_{\beta)} + u^\mu\fpt{}_{\mu  (\alpha}{u}_{\beta)} ) + \Pb \fpt{}_{\alpha \beta}.
$
On the other hand,
the perturbed Einstein tensor $G^{(1)}_{\alpha\beta}\defi \partial_\pertp G(g_\pertp)_{\alpha\beta}|_{\pertp=0}$ only contains $\{t, \varphi\}$ components.
It is then straightforward to show that
the first order field equations that follow from (\ref{eq:EFEs}), i.e.
$ G^{(1)}_{\alpha\beta}=8\pi T^{(1)}_{\alpha \beta}$, imply that
the first order perturbations of the pressure and density must vanish, i.e. $\Ep=\Pp=0$,
and leave only one equation for $\omega(r,\theta)$, that reads \cite{Hartle1967}
\begin{equation}
\frac{\partial}{\partial r}\left( r^4 j \frac{\partial\omega}{\partial r}\right)+ \frac{r^2 j e ^\lambda}{\sin ^3 \theta}\frac{\partial}{\partial \theta}\left( \sin^ 3 \theta \frac{\partial \omega }{\partial \theta}\right)+4r^3j'(\omega - \Omega) =0.
\label{eq:omega}
\end{equation}

The equation for the exterior vacuum region ($-$) is recovered by just setting $j=1$
in the above.

Given the regularity condition at the origin,
the asymptotic behaviour at infinity 
and the matching conditions (\ref{omega_matching})-(\ref{omegap_matching})
the functions $\omega^\pm(r_\pm,\theta_\pm)$ can be shown to be functions
of the corresponding radial coordinates only (see \cite{ReinaVera_ERE2012}).
This is in agreement with Hartle's argument in \cite{Hartle1967}.
In particular, the exterior solution that vanishes at infinity is thus
\begin{equation}
\om=\frac{2J}{\rrm^3}
\label{eq:omega_exterior}
\end{equation}
for some constant $J$ \cite{Hartle1967}.
For later use, it is easy to show that taking into account that $[j]=0$, so that $j(\ro)=1$,
and $[j']=-1/2[\lambda']$ by construction, the difference of equation
(\ref{eq:omega}) yields
\begin{equation}
[\omega'']=[\lambda']\left(\frac{1}{2}\omega'(\ro)
+\frac{2}{\ro}(\omega^+(\ro) -\Omega)\right).
\label{eq:omegapp}
\end{equation}

Regarding the perturbation of the hypersurface, 
let us first note that equations (\ref{Bp}) and (\ref{nupp}) imply that the differences
$[\lambda']$ and $[\nu']$ are proportional to the difference $[\Eb]$.
The remark made after Proposition \ref{teo:first_order_matching} can be now stated in terms
of a physical property of the interior and exterior background configuration:
whenever there is a jump in the energy density at the surface, $Q_1^-(=Q_1^+)$ must vanish
necessarily by (\ref{Q1_matching}).
However, if $[\Eb]=0$ the function $Q_1(\tau,\vartheta)$ is not determined, in principle,
and enters the second order.
Nevertheless, as shown in Section \ref{sec:second_order_matching}
when analysing the determination of the surface of the rotating star at second order,
$Q_1$ will necessarily vanish if $[\Eb']\neq 0$. 
In Appendix \ref{sec:Rgauge} the whole case $[\Eb]=0$ is discussed.

\subsubsection{On gauges at first order}
\label{sec:gauges_first_order}
We discuss next the meaning of the constant $b_1$, how it is related with gauges,
and its role on the determination of the rotation of the perfect fluid star.
Consider a spacetime gauge change in either $(\mathcal{V}_0^\pm, \gback^\pm)$
defined by $\vec{s}_1 = C t \partial_\varphi$ (we drop the $\pm$ for
clarity, the two $C^\pm$ being independent).
The rules of transformation of the first order metric perturbation tensor (\ref{eq:Ksgauged}),
the energy momentum tensor
$T^{(1)g}=T^{(1)}+\mathcal{L}_{\vec s_1}T,$ and of the first order
deformation vector (\ref{eq:Z_gauges}) imply, respectively,
$\omega^{g}=\omega - C$, $\Omega^{g}=\Omega-C$ and
$b_1^{g}=b_1 - C$. First, note that $\omega^+-\Omega$
is independent with respect to that gauge. This quantity
is essentially the $\tilde \omega$ (up to a sign) defined in \cite{Hartle1967}.

As discussed, the first order matching conditions are invariant
under such spacetime gauges (at either or both sides, with
corresponding $C^+$ and $C^-$), that is, the first
order matching conditions (\ref{omega_matching}), (\ref{omegap_matching}),
(\ref{T1_matching}) and (\ref{Q1_matching})
transform to just the same expressions with ${}^{g}$ superscripts.

This first order gauge at either side $\pm$
is fixed (and completely fixed) once the value of the function $\omega^\pm$
is fixed at some point (or infinity).
The equation for $\omega^-$ is usually integrated in the exterior vacuum region
assuming that $\omega^-$ vanishes at $\rrm\to \infty$.
By doing that
$\partial_{t^-}$ is chosen to represent the ``right'' observer
at infinity. At infinity, the vector $\partial_{t^-}$ is thus assumed to be both unit
and orthogonal, with respect to $g_\pertp$ to second order,
to the axial Killing vector $\partial_{\varphi^-}$.
The exterior choice of gauge thus fixes $\omega^-$, and it is given by (\ref{eq:omega_exterior}).

Regarding the interior region, the above spacetime gauge for some $C^+$  can
then be used to get rid of one of the two constants that describe the configuration
at first order, either $b_1$ or $\Omega$, but clearly not both.
The transformations of $b_1$ and $\Omega$ suggest
building a quantity defined on $\Sigma_0$ as 
\begin{equation}
\Omega_{\infty} = \Omega - b_1,
\end{equation}
invariant under the gauge $\vec s_1$.
The meaning of this constant is the following. $\Omega$ defines the rotation
of the fluid flow with respect to the interior observer $\partial_{t^+}$, and
$b_1$ determines the tilt on $\Sigma_0$ between that interior observer $\partial_{t^+}$ and the (already fixed)
exterior observer $\partial_{t^-}$, explicitly
$\partial_{t^+}|_{\Sigma_0}=\partial_{t^-}|_{\Sigma_0}-\pertp b_1\partial_{\varphi}|_{\Sigma_0}$.
The difference $\Omega_{\infty}$ thus describes the
tilt of the fluid flow with respect to the continuous extension of the exterior
observer to the interior, and thence, measures the rotation of the fluid
with respect to the unit non-rotating observer at infinity.

The value of the ``invariant'' quantity $\tilde\omega(r)\defi\omega^+(r) -\Omega$ at the boundary
can then be expressed thanks to the condition (\ref{omega_matching}) as
$\omega^+(\ro)-\Omega=\omega^-(\ro)-\Omega_\infty$,
i.e.  $$\tilde\omega^+(\ro)=2J/\ro^3-\Omega_\infty.$$
This yields the desired
relation between the value of the interior $\tilde\omega^+(\ro)$, integrated via (\ref{eq:omega})
from the origin, $J$ and
the rotation of the star, thus described by $\Omega_\infty$.

In \cite{Hartle1967} the function $\omega$ is assumed to be ``continuous'' by
construction. In the present general setting
that corresponds to a choice of gauge in the interior region for which
$b_1=0$, and therefore $\Omega(=\Omega_\infty)$ corresponds indeed 
to the rotation of the fluid as measured by the unit exterior observer.
The relation between $\Omega$ and  $\OH$ is thus explicitly given
by $\OH{} =\pertp\Omega_{\infty}$. 

In contrast, in \cite{Bradley_etal2007} the gauge in the interior
is chosen so that the interior observer $\partial_{t^+}$ moves with
the fluid, i.e. $\Omega=0$ (comoving gauge).
Thereby, since the freedom one may have in the interior driven by $\vec s_1$
has been already fixed, the price to pay is a rotation in the matching
hypersurface given by the constant $b_1$, which corresponds to the parameter
$-c_4\Omega$  in \cite{Bradley_etal2007}, so that $\Omega_{\infty}$ corresponds to ``$c_4\Omega$'' there.

\subsection{Second order problem}
\label{sec:second_order_problem}
We explore now the second order field equations for a perfect fluid with barotropic equation of state
(the interior) and for the exterior vacuum, and particularise the second order perturbed matching conditions
of Proposition \ref{teo:second_order_matching}.
The conditions on the fluid flow (\ref{eq:conv_rigid}) together with the normalisation condition
now lead to $\vec{u}^{(2)}=\vec{u}^{(2)}{}^{t}\partial_t$, where
$
\vec{u}^{(2)}{}^{t}=e^{-3\nu/2}\left\{\Omega^2 g_{\varphi\varphi}+2\Omega \fpt{}_{t\varphi} +\spt{}_{tt}/2 \right\}.
$

Taking the second $\pertp$-derivative of (\ref{em-family}), evaluating on $\pertp=0$,
and using $\Ep=\Pp=0$,
the second order energy momentum tensor is found to take the form 
\begin{eqnarray}
T^{(2)}_{\alpha \beta} &=& (\Epp+\Ppp){u}_\alpha {u}_\beta
+\Ppp \gback_{\alpha \beta} + \Pb \spt{}_{\alpha \beta} + 2(\Eb+\Pb)\left( u^{(2)}_{(\alpha}{u}_{\beta)} + u^{(1)}_{(\alpha}{u}^{(1)}_{\beta)}\right. \nonumber\\
&&\left.   +  u^\mu \spt{}_{\mu (\alpha}u_{\beta)} + 2 u ^{(1)}{}^\mu \fpt{}_{\mu(\alpha}u_{\beta)} + 2 u {}^\mu \fpt{}_{\mu(\alpha}u ^{(1)}_{\beta)}+ u^\mu u^\rho \fpt{}_{\mu \alpha} \fpt{}_{\rho \beta}\right) . \nonumber
\end{eqnarray}
As follows from (\ref{eq:EFEs})
the second order Einstein field equations consist of equating this to
the second order perturbed Einstein tensor, computed from $g_\pertp$ as
$G^{(2)}_{\alpha\beta}\defi \partial_\pertp \partial_\pertp G(g_\pertp)_{\alpha\beta}|_{\pertp=0}$,
that is
\begin{equation}
G^{(2)}_{\alpha\beta}=8\pi T^{(2)}_{\alpha \beta}.
\label{eq:EFEs2}
\end{equation}

Given that the final purpose of the present work
is to analyse the results in \cite{Hartle1967} regarding the matching problem, we assume
the same angular behaviour of the functions of
the second order perturbation tensor
(for both the interior $+$ and the exterior $-$ regions).
This behaviour is argued in \cite{Hartle1967} to follow from
the non-dependency of
the first order function $\omega$ on any angular coordinate
and equatorial symmetry.
The assumption we take on the functions of $\spt^\pm$ thus reads explicitly
\begin{eqnarray}
h(r,\theta) &=& \hk(r) + \hkt(r) P_2(\cos \theta),\nonumber\\
m(r,\theta) &=& \mk(r) + \mkt(r) P_2(\cos \theta),\nonumber\\
k(r,\theta) &=& \kk(r) + \kkt(r) P_2(\cos \theta),\nonumber\\
\fcross(r,\theta) &=& \fr(r) P_2(\cos \theta), \label{eqs:angular_expansions}
\end{eqnarray}
for both the interior $+$ and the exterior $-$.
A straightforward calculation shows that
the above angular structure assumed on the functions in $\spt$ is inherited, via the
field equations (\ref{eq:EFEs2}), by the second order energy momentum tensor, 
so that
\begin{equation}
\Epp(r,\theta)=\Eppz(r)+\Eppt(r) P_2(\cos\theta),\quad \Ppp(r,\theta)=\Pppz(r)+\Pppt(r) P_2(\cos\theta).
\label{eq:EP_decomp}
\end{equation}

Given that $\Ep=\Pp=0$ the barotropic character of the equation of state to second order (\ref{eq:2ndbarotropic})
translates onto the condition
\begin{equation}
\Epp \Pb'-\Ppp \Eb'=0.\label{baroEOS}
\end{equation}

In order to write down the second order field equations in a
convenient and compact form, let us first
define the following auxiliary ``tilded'' functions 
\begin{eqnarray}
&&\Hzero := \hk -\frac{1}{2} r \nu ' \kk ,\quad \Mzero := \mk -e^{-\lambda/2}\left(e^{\lambda/2}r\kk\right)',\label{HMKzero} \\
&&\Htwo := \hkt - \frac{1}{2} r  \nu ' \fr, \quad \Mtwo := \mkt -e^{-\lambda/2}\left(e^{\lambda/2}r\fr\right)',
\quad \Ktwo := \kkt -\fr.
\label{HMKtwo}
\end{eqnarray}
Clearly these quantities are invariant under the `radial' gauges class of
transformations (\ref{eq:K2_s2}) 
since e.g. both $ h-\frac{1}{2}r\nu'k$ and $ h-\frac{1}{2}r\nu'\fcross$ are.

We introduce now the above decomposed expressions of the relevant quantities
into the field equations (\ref{eq:EFEs2}). By construction the complete set of equations
gets decomposed into the $l=0$ and $l=2$ sectors, which are independent
and can thus be considered separately.
Our purpose in the next two subsections
is to recover and write down the field equations as closely as possible to the
expressions presented in Sections VII and VIII in \cite{Hartle1967}. The explicit correspondences
are made in Section \ref{sec:conclusion}.  

\subsubsection{The EFEs in the $l=0$ sector}
The $l=0$ sector of the field equations (\ref{eq:EFEs2})
can be shown to provide the following expressions
for the second order energy density and pressure\footnote{These two equations correspond to (93) and (94) in \cite{Hartle1967}. Note that a global 2 factor on the right hand side here comes from the definitions (\ref{Energy_epsilon}) and (\ref{Pressure_epsilon})
as compared with the definition of $\Delta G$ in \cite{Hartle1967}, which already contains the $\pertp^2$ and $1/2$ factors.}
\begin{eqnarray}
8\pi\Eppz &=& \frac{4}{r^2}\left(r e^{-\lambda}\Mzero \right)'+\frac{8}{3}r jj'\topert^2 -\frac{1}{3}j^2r^2\oppert^2+ 16\pi r\Eb'\kk, \label{eq:E20} \\
8\pi\Pppz &=&  \frac{4}{r^2 }  \left\{e^{-\lambda} r \Hzero'
-\Mzero \left(8\pi r^2 \Pb+1\right) \right\} + \frac{1}{3} r^2 j^2 \oppert^2 +16\pi r \Pb'\kk, \label{eq:p20}
\end{eqnarray}
plus an equation for $\Hzero''$ of the form $\Hzero''=F_1(\Hzero',\Mzero',\Mzero)$.
A convenient auxiliary definition of the second order pressure
is given by
\begin{equation}
\Ptsppz\defi\frac{\Pppz-2r\Pb'\kk}{2(\Eb+\Pb)}=\frac{\Pppz}{2(\Eb+\Pb)}+\frac{\kk}{r-2M(r)}\left(M(r)+4\pi r^3\Pb\right), \label{eq:P0gorda}
\end{equation}
where (\ref{eq:Pprime}) has been used in the equality. This function is well defined 
at points where $\Eb+\Pb=0$ (see below),
and corresponds to the ($l=0$ part of the) ``pressure perturbation factor'' 
as defined in equation (87) in \cite{Hartle1967}.

On the other hand,
the $l=0$ part of equation (\ref{baroEOS}), i.e. $\Eppz \Pb'-\Pppz \Eb'=0$,
combined with (\ref{eq:E20}),
yields a direct relation between $\Pppz$ and $\mk'$,
which written in terms of ``tilded'' quantities reads
\begin{equation}
  \label{eq:m0prime}
\left(r e^{-\lambda}\Mzero \right)'
=4\pi r^2 \frac{\Eb'}{\Pb'}(\Eb+\Pb)\Ptsppz +\frac{1}{12}j^2r^4\oppert^2-\frac{2}{3}r^3 jj'\topert^2.
\end{equation}

Now, the aforementioned
equation for $\Hzero''$ can be rewritten, using (\ref{eq:p20})
and (\ref{eq:m0prime}) --and (\ref{eq:Mass}) for cosmetics--,
as a first order ODE for $\Ptsppz$,  that reads

\begin{eqnarray}
\Ptsppz{}' &=& -\frac{4\pi (\Eb+\Pb)r^2}{r-2 M(r)}\Ptsppz 
- \frac{(r e^{-\lambda}\Mzero)r^2 }{(r-2M(r))^2}\left( 8\pi \Pb + \frac{1}{r^2}\right) \nonumber \\
&& + \frac{r^4 j^2}{12(r-2M(r))}\oppert^2 + \frac{1}{3} \left(\frac{r^3j^2 \topert^2}{r-2M(r)} \right)'.\label{eq:cons_eq_0}
\end{eqnarray}

The set of functions that determines the $l=0$ sector can thus be taken to be
$\{\Ptsppz,\Mzero\}$, which satisfies the system (\ref{eq:m0prime}), (\ref{eq:cons_eq_0})
given regularity conditions at the origin $r=0$.
Equation (\ref{eq:p20}) can be rewritten as
\begin{equation}
\Hzero'
-re^{\lambda}\Mzero
\left(8\pi \Pb+\frac{1}{r^2}\right) = 4\pi re^{\lambda}(\Eb+\Pb)\Ptsppz- \frac{1}{12} e^{\lambda} r^3 j^2 \oppert^2.
\label{eq:h0prime}
\end{equation}
It is now trivial to check that (see (90) in \cite{Hartle1967})
\begin{equation}
\Ptsppz+\Hzero-\frac{1}{3}r^2e^{-\nu}\topert^2=\gamma
\label{eq:1st_integral_l0}
\end{equation}
for some constant $\gamma$ is a first integral of (\ref{eq:cons_eq_0}) and (\ref{eq:h0prime}).
This relation shows, in particular, that $\Ptsppz$ is well defined in $\rrp\in [0,\ro]$.
The constant $\gamma$ is identified in \cite{Hartle1967} as the second order to background
ratio of the constant injection energy.
In analogy with the Newtonian potential, $\Hzero$ (and thus $\hk$)
is determined up to an arbitrary additive constant.
This constant will be determined once a condition at infinity plus some continuity across the boundary of the body
are imposed. We will discuss that below. Once that is fixed,
the value of $\gamma$ still depends on one factor, that is,
the conditions one may impose on $\Ptsppz$ at the origin.
The latter depends on how one sets the value of the pressure (and thus
of the energy density) of the rotating configuration at the origin with respect to that of
the static configuration. In the present work the pressure is taken to be unchanged after the perturbation,
so that $\Ptsppz(0)=0$, as in \cite{Hartle1967} and subsequent works.
Leaving $\Ptsppz(0)$ as an extra parameter of the model does not introduce any remarkable effect
for the purposes of our analysis. We refer to \cite{Bradley_etal2007} (Section III A.) for a deeper discussion.

The fact that $\kk$ is ``pure gauge'' translates onto the fact that it does not enter
the set of equations, and it is therefore not determined.
The quantitites $\Eppz$ and $ \Pppz$ are gauge dependent, and
can only be computed, from (\ref{eq:E20}) and (\ref{eq:p20})
respectively, once $\kk$ is specified, i.e. by fixing the `radial' gauge.
The quantities independent of that choice, and thus the relevant ones, correspond to
$\Eppz-2\Eb'r\kk$ and $\Pppz-2\Pb'r\kk$. This is the motivation
for the introduction of the auxiliary  function $\Ptsppz$.

The equations for $\{\Hzero^-,\Mzero^-\}$ in the \textbf{vacuum exterior} are obtained by using
(\ref{eqs_back_vacuum}) and the first order solution (\ref{eq:omega_exterior}) in equations
(\ref{eq:E20}), (\ref{eq:p20}) with their left hand sides and $\Pb$ and $\Eb$ set to zero.
The boundary condition is set so that $\Hzero^-$ and $\Mzero^-$ vanish at infinity $\rrm\to \infty$.
The solutions vanishing at infinity are thus given by
\begin{eqnarray}
  &&r_- e^{-\lambda(r_-)}\Mzero^- (\rrm)=  \delta M - \frac{J^2}{\rrm^3},\label{eq:m_0_ext}\\
  &&\Hzero^- (\rrm)= -\frac{\delta M}{\rrm-2M}+\frac{J^2}{\rrm^3(\rrm -2M)},\label{eq:h_0_ext}
\end{eqnarray}
where $\delta M$ is an arbitrary constant.
As mentioned above, the function $\kk^-$ remains undetermined, under the
condition that is also vanishes at infinity.

\subsubsection{The EFEs in the $l=2$ sector}
\label{sec:efes_l2}
Apart from the two field equations that provide the energy density and pressure,
the $l=2$ sector provides three equations. The whole set of equations
can be shown to be equivalent to the system

\begin{eqnarray}
   \Htwo'+\Ktwo'&=&-\nu'\Htwo +\left(\frac{1}{r}+\frac{\nu'}{2}\right)
  \left(-\frac{2}{3} r^3 j j'\topert^2 +\frac{1}{6}j^2r^4\oppert^2 \right)\label{eq:l20},\\
   \Htwo'&=&\left\{-\nu'+\frac{r}{(r-2M(r)) \nu'}
    \left(8\pi(\Eb+\Pb)-\frac{4M(r)}{r^3}\right)\right\}\Htwo
  -\frac{4(\Htwo+\Ktwo)}{r\nu'(r-2M(r))}\nonumber\\
  && +\frac{1}{6}\left(\frac{1}{2}r\nu'-\frac{1}{(r-2M(r)) \nu'}\right)r^3j^2\oppert^2 \nonumber\\
  &&  -\frac{1}{3}\left(\frac{1}{2}r\nu'+\frac{1}{(r-2M(r)) \nu'}\right)r^2(j^2)'\topert^2\qquad \label{eq:l21}
\end{eqnarray}
plus the equation
\begin{equation}
 \Mtwo =\frac{1}{6}r^4j^2\oppert^2-\frac{1}{3}r^3(j^2)'\topert^2-\Htwo.\label{eq:mt2}
\end{equation}
The expressions for the energy density and pressure can then be written as
\begin{eqnarray}
\Eppt -2\Eb'r\fr&=& \frac{4  {\Eb}' }{3\nu' }\left(3 \Htwo +e^{-\nu}r^2 \topert^2\right),\label{E22}\\
\Pppt -2\Pb'r\fr&=& -\frac{2}{3} (\Eb+\Pb) \left(3 \Htwo +e^{-\nu} r^2  \topert^2\right).\label{P22}
\end{eqnarray}
Note that we have kept the background function $\nu'$ explicitly
in order to ease the eventual comparison with the expressions in \cite{Hartle1967}.
In the $l=2$ sector imposing the condition of a barotropic EOS does not add any further condition.  

The convenient ``pressure perturbation factor'' in this case corresponds to the following definition
\begin{equation}
  \label{eq:P2gorda}
  \Ptsppt\defi\frac{\Pppt-2\Pb'r\fr}{2(\Eb+\Pb)}, 
\end{equation}
so that (\ref{P22}) just reads
\begin{equation}
\label{eq:integral_l2}
\Ptsppt+\Htwo+\frac{1}{3}e^{-\nu} r^2  \topert^2=0.
\end{equation}
This corresponds to (91) in \cite{Hartle1967}, and, together with the above (\ref{eq:1st_integral_l0})
form the $l=0$ and $l=2$ parts of the first integral $\gamma$, (86) in \cite{Hartle1967}.

The interior region is thus determined by the solution
of the pair  $\{\Htwo^+,\Ktwo^+\}$ to the system (\ref{eq:l20}), (\ref{eq:l21})
given regularity conditions at the origin $\rrp\to 0$,
up to an arbitrary constant, say $A'$.
Then, $\Mtwo^+$ is directly obtained from (\ref{eq:mt2}).
The function $\fr(r)$ does not enter the equations, and thus it is, as expected,
pure gauge.

In the \textbf{vacuum exterior} region only equations (\ref{eq:l20})-(\ref{eq:mt2})
apply. Using (\ref{eqs_back_vacuum}), so that in particular $\Pb=0$,
and (\ref{eq:omega_exterior}), and
given the asymptotic behaviour at $\rrm\to \infty$,
the whole set of exterior functions $\{\Htwo^-,\Ktwo^-,\Mtwo^-\}$ is
integrated and read
\begin{eqnarray}
\Htwo^- &=&A Q_2^2\left(\frac{\rrm}{M}-1 \right) + \frac{J^2}{\rrm^3}\left( \frac{1}{M} + \frac{1}{\rrm}\right),
\label{h2_vacuum}\\
\Ktwo^-+ \Htwo^-&=&A \left\lbrace \frac{2M}{\sqrt{\rrm(\rrm-2M)}}Q_2^1\left(\frac{\rrm}{M}-1   \right)\right \rbrace - \frac{J^2}{\rrm^4},\label{k2_vacuum}\\
\Mtwo^- &=& -AQ_2^2\left(\frac{\rrm}{M}-1 \right) + \frac{J^2}{\rrm^3}\left( \frac{1}{M} - \frac{5}{\rrm}\right),\label{m2_vacuum}
\end{eqnarray}
where $Q^m_l(x)$ stand  for the associated Legendre functions of the second kind, and $A$ is an arbitrary constant.
The constants $A'$ and $A$ are to be determined once the relations between $\{\Htwo^+,\Ktwo^+\}$
and $\{\Htwo^-,\Ktwo^-\}$ on the matching hypersurface $\Supfamp_0$ are determined (see below).

\subsubsection{The matching of the second order problem}
\label{sec:second_order_problem_matching}
We particularize first the matching conditions as given in Proposition
\ref{teo:second_order_matching} for the particular angular expansion
of the perturbation functions (\ref{eqs:angular_expansions}) at both
sides.  The field equations in the background allow us to express the
differences $[\lambda']$ and $[\nu'']$ in terms of $[\Eb]$ by direct
use of (\ref{Bp}) and (\ref{nupp}).
However, we will not use those
relations in some places, nor the explicit form of $\nu_-(r_-)$ in the
exterior, to keep more compact expressions.
Let us recall that condition (\ref{condition_f}) now just reads $M\neq0$
given the exterior is vacuum.
Clearly, for all pairs $f^\pm(r_\pm,\theta_\pm)$ such that
$f=f_0(r)+f_2(r)P_2(\cos\theta)$ we have
$[f]=[f_0]+[f_2]P_2(\cos\vartheta)$. Note that $[f_0]$ and $[f_2]$ are
constants. 

Equation (\ref{eq:mc:k}) is thus satisfied if and only if $c_1=0$ plus
\begin{equation}
  \difkt=\diffr\label{eq:k2_matching}.
\end{equation}
The constant $c_2$ just corresponds to the difference $\difkz$, i.e. $\difkz=c_2$.

Likewise, equation (\ref{eq:mc:h}) is satisfied if and only if $H_1=0$ plus
\begin{eqnarray}
  &&\difhz=\frac{H_0}{2}+\frac{1}{2}\ro\nu'(\ro)\difkz\label{eq:h0_matching}\\
  &&\difht=\frac{1}{2}\ro \nu'(\ro)\diffr\label{eq:h2_matching}.
\end{eqnarray}
Equation (\ref{eq:mc:Q2}), since $c_1$ and $H_1$ must vanish,
imposes a very particular expansion of
$[\Qt2](\vartheta)$, explicitly
\[
[\Qt2](\vartheta)=[\Qttwoz]+[\Qttwot] P_2(\cos\vartheta)
\]
for some constants $[\Qttwoz]$ and $[\Qttwot]$.
Equation (\ref{eq:mc:Q2}) is thus equivalent to the pair
\begin{eqnarray}
  &&[ \Qttwoz ]=2\ro e^{-\nu(\ro)/2}\difkz \label{eq:qt20_matching}\\
  &&[ \Qttwot ]= 2\ro e^{-\nu(\ro)/2}\diffr, \label{eq:qt22_matching}
\end{eqnarray}
where here, and in the following expressions, equation (\ref{eq:cont_back}) is used to set
$\lambda(\ro)=-\nu(\ro)$.
Take now the equations for the differences $\difm$ and $\difhp$.
In the case $[\Eb]\neq 0$ ($[\lambda']\neq 0$ and $[\nu'']\neq 0$),
for which $Q_1=0$ necessarily, we recall we necessarily have $\Qt2^+=\Qt2^+(\vartheta)$
and therefore both $\Qt2^\pm$ due to the above, so that
\begin{equation}
\Qt2^\pm(\vartheta)=\Qttwoz^\pm+\Qttwot^\pm P_2(\cos\vartheta),
\label{eq:Q2_decomp}
\end{equation}
with constants $\Qttwoz^\pm$ and $\Qttwot^\pm$.
Thence, equation (\ref{eq:mc:m}) holds iff
\begin{eqnarray}
  &&\difmz = \ro \difkpz +\frac{1}{4}e^{\nu(\ro)/2}[\lambda']\Qttwoz^++\frac{1}{2}(\ro\lambda_-'(a)+2)\difkz \label{eq:m0_matching}\\
  &&\difmt = \ro \difkpt +\frac{1}{4}e^{\nu(\ro)/2}[\lambda']\Qttwot^++\frac{1}{2}(\ro\lambda_-'(a)+2)\diffr,\label{eq:m2_matching}
\end{eqnarray}
while equation (\ref{eq:mc:hp}) does whenever
\begin{eqnarray}
  &&\difhpz = \frac{1}{2}\ro\nu'(\ro) \difkpz +\frac{1}{4}e^{\nu(\ro)/2}[\nu'']\Qttwoz^+
  +\frac{1}{2}(\ro\nu''_-(a)+\nu'(\ro))\difkz, \label{eq:hp0_matching}\\
  &&\difhpt = \frac{1}{2}\ro\nu'(\ro) \difkpt +\frac{1}{4}e^{\nu(\ro)/2}[\nu'']\Qttwot^+
  +\frac{1}{2}(\ro\nu''_-(a)+\nu'(\ro))\diffr.\label{eq:hp2_matching}
\end{eqnarray}

In the case $[\Eb]=0$, the equation $[\omega'']Q_1=0$ provides no information,
since $[\omega'']=0$ as follows from (\ref{eq:omegapp}) and (\ref{Bp}).
On the other hand, the equations corresponding to
(\ref{eq:mc:m}) and (\ref{eq:mc:hp}) with the changed terms
(\ref{eq:changes_for_Q1}) contain a term proportional to
$[\Eb'](Q_1)^{2}$. If $[\Eb']=0$ we recover the above equations (with $[\lambda']=[\nu']=[\Eb]=0$)
and therefore one only needs considering the case $[\Eb']\neq 0$.
In that case the equations imply, analogously,
that $Q_1$ does not depend on $\tau$ and that it
must satisfy $(Q_1)^{2}=q_0+q_2P_2(\cos\vartheta)$ for some constants $q_0$ and $q_2$.

Some remarks are in order now, which will lead us eventually to the determination of the deformation
of the matching hypersurface at second order in any `radial' gauge --recall that the deformation vectors $\vec Z$ are gauge dependent,
and therefore the functions $Q$ describe the deformation with respect to the gauge chosen.
The appropriate quantities are constructed as follows
\begin{equation}
\defor_0\defi \Qttwoz-2ae^{-\nu(\ro)/2}\kk(\ro),\qquad
\defor_2\defi \Qttwot-2ae^{-\nu(\ro)/2}\fr(\ro)
\label{eq:def_deformation}
\end{equation}
on $\Sigma_0$ from either side $+$ and $-$. 
These two quantities
are `radial'-gauge independent, since the gauge defined by
$\vec V_2=2S(r,\theta)\partial_r$ (and $\vec s_1=Ct\partial_\phi$)
induces via (\ref{eq:Z_gauges}) the transformation $\Qt2^g=\Qt2+2Se^{\lambda(\ro)/2}$, while
$k^g=k+S/r$ and $\fcross^g=\fcross+S/r$, see (\ref{eq:K2_s2}).
On the other hand, the relations (\ref{eq:qt20_matching}) and (\ref{eq:qt22_matching}) just read
\begin{equation}
  [ \defor_0]=0,\qquad[ \defor_2]=0,\label{eq:deformation_matching}
\end{equation}
meaning  that the quantities coincide as computed from either side.
How the actual deformation $\Supfamp^+_\pertp$ out from the spherical $\Supfamp_0$ is encoded in terms of $\defor_0$ and $\defor_2$
is described in Appendix \ref{sec:Rgauge}.

The above matching conditions to second order have yet to be combined with the constraints provided by
the field equations at either side.
We obtain the final expressions of the matching conditions to second order
using the second order field equations for the perfect fluid interior and the vacuum exterior next.

Regarding the $l=0$ sector, the differences of the field equations do not provide
any constraints to the matching conditions in the sense that the differences $\difkz$ and $\difkpz$ remain
arbitrary (constants). This, as expected, is related to the fact that $\kk$ is pure gauge.
The $l=0$ matching conditions (\ref{eq:h0_matching}), (\ref{eq:m0_matching})
and (\ref{eq:hp0_matching}) can be written in terms of the ``tilded'' functions
(\ref{HMKzero}) and the deformation functions (\ref{eq:def_deformation})
in the case $[\Eb]\neq 0$ as follows,
\begin{eqnarray}
  &&[\Hzero]=\frac{H_0}{2},\label{eq:ht0_matching}\\
  &&[\Hzero']=\frac{\ro-M}{\ro(\ro-2M)}[\Mzero],
          \label{eq:htp0_matching}\\
  &&[\Mzero]=2\pi[\Eb] e^{-\nu(\ro)/2}\ro\defor_0,\label{eq:mt0_matching} 
\end{eqnarray}
while if
$[\Eb]= 0$
equation (\ref{eq:mt0_matching}) is replaced by
\begin{equation}
[\Mzero]= - 2\pi[\Eb'] e^{-\nu(\ro)/2}\ro q_0^2.
\label{eq:mt0_matching_E0}
\end{equation}
The background matching configuration relations
(\ref{Bp}) and (\ref{nupp}) have been used to express the background difference functions in terms
of $[\Eb]$, which is just $\Eb_+(a)$ (vacuum exterior),
together with (\ref{eqs_back_vacuum}) to write
\begin{equation}
\ro\nu'(\ro)=\frac{2M}{a-2M}=e^{-\nu(\ro)}\frac{2M}{\ro}.
\label{eq:anup}
\end{equation}
The arbitrariness in shifting $\Hzero^+(\rrp)$ corresponds here to the appearance of the free constant $H_0$.
One can always fix the shift in $\Hzero^+(\rrp)$ in the interior simply by choosing $H_0$.
This just mirrors the fact that in Newtonian theory
the potential is fixed at infinity and then taken to the interior of the body simply by imposing continuity across the boundary.

It must stressed, however, that the argument about the ``continuity'' of $\Hzero$
does not stand for the other function $\Mzero$ in general. Consider first the difference of equation (\ref{eq:p20})
for a vacuum exterior combined with the two matching
conditions (\ref{eq:ht0_matching}), (\ref{eq:htp0_matching}) at hand, which leads to the relation
\begin{equation}
  \label{eq:mt0_matching2}
  [\Mzero]=-4\pi\frac{\ro^3}{M}[\Eb] \Ptsppz(\ro),
\end{equation}
after using the definition (\ref{eq:P0gorda}). Note that this equation holds always,
irrespective of the vanishing (or not) of $[\Eb]$.
Now, in the case $[\Eb]\neq 0$,  (\ref{eq:mt0_matching}) can be finally rewritten as
\begin{equation}
\left(2[\Eb]\Ptsppz(\ro)=\right)\quad \Pppz{}(\ro)-2\ro\Pb'(\ro)\kk^+(\ro) =-  \frac{M}{\ro^2}e^{-\nu(\ro)/2} [\Eb]\defor_0.
\label{eq:Q0p0}
\end{equation}
In the $[\Eb]= 0$ case equation (\ref{eq:mt0_matching2}) clearly implies $[\Mzero]=0$ and therefore
(\ref{eq:mt0_matching_E0}) yields $$[\Eb']q_0= 0.$$

The implication of (\ref{eq:mt0_matching2}) is that
\emph{the values of the functions $\Mzero^+(\ro)$ and $\Mzero^-(\ro)$
coincide if and only if $[\Eb] \Ptsppz(\ro)= 0$.} This fact turns out to be in contradiction
with the assumption made in \cite{Hartle1967} stating that $\mH_0$ is ``continuous'' at the boundary,
with consequences on the determination of $\delta M$.
We devote the concluding section to analyse this discrepancy and provide the correct
expression for $\delta M$ .

Finally, the field equation (\ref{eq:p20}) at both sides ($\pm$) can be used to replace
the condition (\ref{eq:htp0_matching}) by (\ref{eq:Q0p0}).
To sum up, given the Einstein's field equations hold, in the $l=0$ sector the set of matching conditions
can be given by the two conditions (\ref{eq:ht0_matching}) and
either (\ref{eq:mt0_matching}) or  (\ref{eq:mt0_matching2}),
plus the relation (\ref{eq:Q0p0}).

In the $l=2$ sector things are different, in the sense that the field equations provide, in principle, further constraints
to the matching conditions.
Taking the differences of the field equations (\ref{eq:l20}), (\ref{eq:l21}) and (\ref{eq:mt2})
we obtain three equations for the differences $[\Mtwo]$, $[\Ktwo']$, $[\Htwo']$
which have to be added to the relations in (\ref{eq:m2_matching}) and (\ref{eq:hp2_matching})
and the relations (\ref{eq:k2_matching}) and (\ref{eq:h2_matching}) that already determine $[\Ktwo]$ and $[\Htwo]$
trivially.
The number of independent equations turns out to be four plus these two trivial ones,
and can be finally cast, when $[\Eb]\neq 0$ ($\Rightarrow Q_1=0$), as 
\begin{eqnarray}
  &&[\Ktwo]=0,\qquad [\Htwo]=0,\label{eq:contl2}\\
  &&[\Eb]\left\{\Htwo(\ro)-\frac{1}{4}\nu'(\ro)e^{\nu(\ro)/2}
    \defor_2+\frac{1}{3}\ro^2 e^{-\nu(\ro)}\left(\frac{2J}{\ro^3}-\Omega_\infty\right)^2\right\}=0,\label{eq:Q2h2}
\end{eqnarray}
plus
\begin{eqnarray}
 &&[\Htwo']=4\pi[\Eb]\frac{\ro^2}{M}\Htwo(\ro)+\frac{4}{3}\pi[\Eb]\frac{\ro^2}{M} 
  e^{-2\nu(\ro)}\left((\ro-M)^2+M^2\right)\left(\frac{2J}{\ro^3}-\Omega_\infty\right)^2,\label{eq:l2_a}\\
  &&[\Ktwo']=-4\pi[\Eb]\frac{\ro^2}{M}\Htwo(\ro)-\frac{4}{3}\pi[\Eb]\frac{\ro^3}{M}(\ro - 2M) 
e^{-\nu(\ro)} \left(\frac{2J}{\ro^3}-\Omega_\infty\right)^2,\label{eq:l2_b}\\
  &&[\Mtwo]=\frac{8}{3}\pi \ro^4[\Eb] e^{-\nu(\ro)} \left(\frac{2J}{\ro^3}-\Omega_\infty\right)^2,\label{eq:l2_c}
\end{eqnarray}
where we have used, in particular, that
\[
[j'\topert^2]=-\frac{1}{2}[\lambda']\left(\frac{2J}{\ro^3}-\Omega_\infty\right)^2=
-4\pi\ro[\Eb]e^{-\nu(\ro)}\left(\frac{2J}{\ro^3}-\Omega_\infty\right)^2
\]
given the exterior region is vacuum.
Therefore, for $[\Eb]\neq 0$ 
the set of matching conditions for the $l=2$ sector is composed by only three equations,
given by the two in (\ref{eq:contl2}) and (\ref{eq:Q2h2}). The three relations (\ref{eq:l2_a}),
(\ref{eq:l2_b}) and (\ref{eq:l2_c}) are now a consequence of (\ref{eq:contl2}) and (\ref{eq:Q2h2})
and the field equations (\ref{eq:l20}), (\ref{eq:l21}) and (\ref{eq:mt2}) at both sides.

Regarding the $[\Eb]=0$ case, all the above equations (\ref{eq:contl2})-(\ref{eq:l2_c}) hold except for (\ref{eq:Q2h2}),
which has to be substituted by $[\Eb']q_2=0$.
We thus have $[\Eb']Q_1=0$.
As a first consequence, the above matching conditions (\ref{eq:contl2})-(\ref{eq:l2_c})
always hold true, irrespective of whether or not $[\Eb]$ vanishes.
Finally, if $[\Eb']\neq 0$ then
\[
Q_1=0.
\]

Let us summarise the main results in this section so far
in the form of the following theorem. 

\begin{theorem} \label{teo:final_matching}
Let $(\mathcal{V}_0,\gback)$ with $\Sigma_0$ be the static and
spherically symmetric background matched spacetime configuration,
perturbed at either side to first order by the functions
$\omega^\pm(r_\pm,\theta_\pm)$ through $\fpt^\pm$ as defined in
(\ref{fopert_tensor})
plus the unknowns $Q_1^\pm(\tau,\vartheta)$ and
$\vec T_1^\pm(\tau,\vartheta)$, as described
in Proposition \ref{teo:first_order_matching}, so that the
first order matching conditions
(\ref{omega_matching}) and (\ref{omegap_matching})
plus (\ref{T1_matching}) and (\ref{Q1_matching}) hold.
Let the configuration be perturbed to second order
by $\spt^\pm$ as defined in (\ref{sopert_tensor}),
plus the unknowns $\Qt2^\pm(\tau,\vartheta)$ and
$\vec{T}^\pm_2(\tau,\vartheta)$ on $\Sigma_0$, and
assume that the interior region ($+$) satisfies the field equations for a
perfect fluid with barotropic equation of state and
that the exterior ($-$) region is asymptotically flat and
satisfies the vacuum field equations up to second order.
The energy density $\Eb(\rrp)$ and pressure $\Pb(\rrp)$ of the
interior background configuration are given by (\ref{eq:lambdaprime})
and (\ref{eq:nuprime}) and must satisfy (\ref{eq:Pprime}).
The background exterior vacuum solution is given by (\ref{eqs_back_vacuum}),
and we assume $M>0$. Consider the convenient background quantities
defined in (\ref{eq:Mass}).

Let  $\vec u_\pertp$ be the unit vector fluid
corresponding to the interior family of metric tensors
$g^+_\pertp = \gback^+ + \pertp \fpt{}^+ + \frac{1}{2}\pertp^2 \spt{}^+ + \mathcal{O}(\pertp^3)$.
Assume that $\vec u_\pertp$ satisfies (\ref{eq:conv_rigid}) for some
constant $\Omega$.
Let $J$ be defined by the first order exterior solution
(\ref{eq:omega_exterior}). 

Assume finally at both sides ($\pm$) that the
first order function $\omega$ depends only on the radial coordinate,
and that the second order functions are decomposed in Legendre
polynomials in terms of $\{\hk,\hkt,\mk,\mkt,\kk,\kkt,\fr\}$ by
(\ref{eqs:angular_expansions}).

Then
\begin{enumerate}
\item The second order pressure $\Ppp$ and
energy density $\Epp$ of the fluid inherit the same angular dependency,
that is, (\ref{eq:EP_decomp}) hold for some
$\Eppz(r),\Eppt(r),\Pppz(r)$ and $\Pppt(r)$. With the help of convenient
alternative ``tilded'' counterparts, defined in
(\ref{HMKzero})-(\ref{HMKtwo}) plus (\ref{eq:P0gorda})
and (\ref{eq:P2gorda}), the Einstein's field equations in the interior
can be expressed as the system (\ref{eq:m0prime}), (\ref{eq:cons_eq_0}) and
(\ref{eq:1st_integral_l0}) for some constant $\gamma$ for
the set $\{\Ptsppz^+,\Mzero^+,\Hzero^+\}$
plus the system (\ref{eq:l20}), (\ref{eq:l21}), (\ref{eq:mt2})
for the set $\{\Htwo^+,\Ktwo^+,\Mtwo^+\}$.
The vacuum solution at second order is given by (\ref{eq:m_0_ext}),
(\ref{eq:h_0_ext}), (\ref{h2_vacuum}), (\ref{k2_vacuum}) and (\ref{m2_vacuum})
where $\delta M$ and $A$ are arbitrary constants.

\item Given the  Einstein's field equations of the previous point are satisfied,
the necessary and sufficient conditions that the metric
perturbation tensors $\spt^\pm$ must
satisfy to fulfil the second order matching conditions 
are given by
(\ref{eq:ht0_matching}) and (\ref{eq:mt0_matching2}) for the sets
$\{\Ptsppz^\pm,\Mzero^\pm,\Hzero^\pm\}$, with arbitrary constant $H_0$,
and the two equations in (\ref{eq:contl2}) for the sets $\{\Htwo^\pm,\Ktwo^\pm,\Mtwo^\pm\}$.

\end{enumerate}\finn
\end{theorem}

Regarding the deformation of the boundary $\Sigma_0$,
expressions (\ref{eq:Q0p0}) and (\ref{eq:Q2h2}) 
show explicitly how
the quantities $\defor_0$ and $\defor_2$, and thus the deformation of $\Sigma_0$,
are linked in a `radial'-gauge invariant manner to the jump
in the pressure at second order across the boundary of the star
through the value of the energy density of the background configuration at $\Sigma_0$.
Whenever $[\Eb]\neq 0$, equations (\ref{eq:Q0p0}) and (\ref{eq:Q2h2}) directly determine 
$\defor_0$ and $\defor_2$
in terms of $\Ptsppz(\ro)$ and $\Htwo(\ro)$ respectively, which are quantities that are obtained by integration
from the origin.
Equations (\ref{eq:Q0p0}) and (\ref{eq:Q2h2}) can then be cast as
\begin{eqnarray}
\defor_0&=&-\frac{2\ro^2}{M}e^{\nu(\ro)/2}\Ptsppz(\ro), \label{eq:defor0}\\
\defor_2&=&e^{-\nu(\ro)/2}
\frac{2\ro(\ro-2M)}{M}\left(\Htwo(\ro)+\frac{1}{3}\frac{\ro^3}{\ro-2M}\left(\frac{2J}{\ro^3}-\Omega_\infty\right)^2\right)\nonumber\\
&=&-\frac{2\ro^2}{M}e^{\nu(\ro)/2}\Ptsppt(\ro),
\label{eq:defor2}
\end{eqnarray}
after using (\ref{eq:anup}) and (\ref{eq:integral_l2}) in the first and second equalities in the latter,
respectively.

However, if $[\Eb]=0$, since  $\Qt2^\pm$ are only defined
on $\Sigma_0$
we cannot 
determine the deformation directly from the above,
in the same way $Q_1$ is undetermined in the first order problem in that case.
This is to be expected.
In fact, as an extreme case,
when matching two vacuum regions the matching hypersurface is not determined in general.
The idea is that in order to have a boundary determined by the matching,
the energy density
must depart from zero as one moves to the interior, so that the star indeed extends no further than,
and up to, that surface. A sufficient condition is that $[\Eb']\neq 0$.
In that case it can be shown that one can make use of the gauge that follows the surfaces of constant energy density, which has been used so
extensively in the literature, specially in \cite{Hartle1967}.
In order to determine the deformation
one can then extend $\defor_0$ and $\defor_2$ to the interior, say using some functions
$\xi_0(\rrp)$ and $\xi_2(\rrp)$ in a convenient way, using that gauge, to finally obtain the deformation by continuity.
This is discussed in Appendix \ref{sec:Rgauge},
where it is shown, in particular, that (\ref{eq:defor0}) and (\ref{eq:defor2}) will hold
also when $\Eb(\ro)=0$, under the condition that 
the  gauge that follows the surfaces of constant energy density exists.
This suggests the fact that equations (\ref{eq:Q0p0}) and (\ref{eq:Q2h2}) are expected to appear again at
higher orders, in the same way the condition $[\Eb]Q_1=0$ of the first order problem appears
as $[\Eb'](Q_1)^2=0$ at second order.

\section{Conclusion: comparing with Hartle's results}
\label{sec:conclusion}
The gauge used in \cite{Hartle1967} at first order corresponds
to setting $b_1=0$ here, while at second order the starting point is the choice of gauge that
corresponds here to setting $\kk^\pm=0$ and $\fr^\pm=0$. We refer to this choice as the $k$-gauge. 
At some point another gauge comoving with the deformation is introduced.
A discussion of the use of that gauge in \cite{Hartle1967} (also in \cite{Bradley_etal2007}) can be found in
Appendix \ref{sec:Rgauge}.

In the $k$-gauge all the ``tilded'' functions (\ref{HMKzero}) and (\ref{HMKtwo}) correspond to
the non-''tilded'' counterparts, and in the interior region ($+$), the functions $\Ptsppz$
and $\Ptsppt$  are just rescalings of their respective $\Pppz$ and $\Pppt$, that is,
$\Ptspp_{0/2}=\Ppp_{0/2}/(2(\Eb+\Pb)):=\Pspp_{0/2}$.
To avoid having to rewrite all the previous equations without tildes we will simply
use the ``tilded'' functions in what follows.

Let us first concentrate on the $l=0$ sector.
Regarding the interior region,
the system (\ref{eq:m0prime})-(\ref{eq:cons_eq_0}) plus equation
(\ref{eq:h0prime}) for the set
$\{re^{-\lambda}\Mzero^+,\Ptspp_0,\Hzero^+\}$,
as functions of $r$ ($\rrp$ in fact)
coincide one by one with the coupled equations
(97) and (100), plus (98) in \cite{Hartle1967} for $\{\mH_0,p^H_0{}^*,\hH_0\}$
as functions of $R$, which has the same range as $r_+$.
To be precise, one can forget about $r_+$ and $R$ and just
establish a common variable $s$, so that the sets of equations here and in
\cite{Hartle1967} hold in the range $s\in (0,\ro]$. Given common 
conditions at $s\to 0$ the problem for $\{re^{-\lambda}\Mzero^+,\Ptspp_0,\Hzero\}$
coincides with the problem for $\{\mH_0,p^H_0{}^*,\hH_0\}$ and therefore
$\mH_0(s)=se^{-\lambda(s)}\Mzero(s)$, $p^H_0{}^*(s)=\Ptspp_0(s)$
and $\hH_0(s)=\Hzero(s)$ (up to a free additive constant) necessarily for $s\in (0,\ro]$, i.e.
\emph{in the interior region}.

In the vacuum exterior region $\mk=\Mzero$ and $\hk=\Hzero$
are given by (\ref{eq:m_0_ext}) and (\ref{eq:h_0_ext}) respectively.
Again, these two expressions correspond to (105) and (106) in \cite{Hartle1967}
for $\mH_0$ and $\hH_0$ respectively, in terms of a variable $r$
in the range $r\in [\ro,\infty)$.

Therefore, the matching conditions for the function $\Hzero$ given by
(\ref{eq:h0_matching}) and (\ref{eq:ht0_matching}), and for the function $\Mzero$
given by (\ref{eq:mt0_matching}), translate directly to matching conditions
on $\hH_0$ and $\mH_0$.
As discussed previously, the free additive constant in $\Hzero^+$ (and so in $\hH_0$) can be used 
to set $H_0=0$. In an abuse of terminology, the assumption of a ``continuous'' $\hH_0$ is thus consistent.

The function $\mH_0$ is also assumed to be ``continuous'' in \cite{Hartle1967} Section VII,
when the value of $\mH_0(a)$ as computed from the interior is equated to the expression of
$\mH_0(r)$ in the exterior at $r=a$ in order to obtain the constant $\delta M$ in (107)\cite{Hartle1967}.
However, the correct matching condition is given by (\ref{eq:mt0_matching2}), which
in the $k$-gauge, and since $[\lambda]=0$, can be expressed as 
\[
[\mH_0]=-4\pi\frac{\ro^3}{M}(a-2M)\Eb(\ro) p^H_0{}^*(\ro)
\]
using the notation in \cite{Hartle1967}. As a result, given the value $\mH_0(a)$ as computed from the interior,
the value of the change in mass in (\ref{eq:m_0_ext}) is given by
\begin{equation}
  \label{eq:deltaM}
  \delta M=\mH_0(a)+\frac{J^2}{\ro^3}+4\pi\frac{\ro^3}{M}(a-2M)\Eb(\ro) p^H_0{}^*(\ro).
\end{equation}
The last term corresponds to the jump of the values of $\Mzero$ at the boundary, 
and it is not present
in the expression for the change of mass in (107)\cite{Hartle1967} and in
the subsequent works, e.g. \cite{HartleThorne1968,HartleThorne1969}. Of course, whenever the density of mass-energy vanishes at the surface
of the star this term has no consequences. This will happen in many situations, as in the cases
of equations of state that imply the vanishing of the energy density at points
where the pressure vanishes, polytropes for instance.
In fact, in the series of papers started by \cite{HartleThorne1968,HartleThorne1969}
all the equations of state considered satisfy that condition, and therefore the computation of the change of mass
is not affected by the correction in (\ref{eq:deltaM}).

However, in more general situations that is not going to be the case. As an example, 
models for quark stars
that rely on a non-zero value of $E$ at the surface 
have been considered in the literature (see e.g. \cite{ColpiMiller}).
In particular, models of stars based on a constant background $E$ in the interior are affected
by that term and the computation of the change in mass should be corrected.
A future work will we devoted to revisit the models presented in \cite{Chandra_Miller1974}, and find
numerically the discrepancy in the values of $\delta M$.

Let us now jump to the $l=2$ sector. In the interior region
the equation (\ref{eq:mt2}) plus
the system (\ref{eq:l20})-(\ref{eq:l21}) for the set
$\{re^{-\lambda}\Mtwo^+,\Htwo^++\Ktwo^+,\Htwo^+\}$
as functions of $\rrp$ 
coincide one by one with equation (120) plus the coupled equations
(125)-(126) in \cite{Hartle1967} for $\{\mH_2,v^H:=\hH_2+\kH_2,\hH_2\}$
as functions of $R$, which has the same range as $r_+$.
The same argument as in the $l=0$ sector shows that the problems
coincide and therefore we can set $\mH_2(s)=s e^{-\lambda(s)}\Mtwo(s)$, $\hH_2(s)=\Htwo(s)$
and $\kH_2(s)=\Ktwo(s)$ for $s\in (0,\ro]$.
In the vacuum exterior region $\hkt=\Htwo$ and  $\kkt=\Ktwo$ are given by (\ref{h2_vacuum}) and (\ref{k2_vacuum}), 
which correspond to (139) and (140) in \cite{Hartle1967} respectively
in terms of a variable $r$ in the range $r\in [\ro,\infty)$.
The comparison of (\ref{eq:integral_l2}) with (91) in \cite{Hartle1967} implies the correspondence $p^H_2{}^*(s)=\Ptspp_2(s)$.
The two matching conditions 
in (\ref{eq:contl2}) simply state that 
$\hH_2$ and $\kH_2$ are ``continuous''
on the boundary. The assumption made in \cite{Hartle1967} regarding the $l=2$ sector is thus consistent.
This ``continuity'' of $\hH_2$ and $\kH_2$ is finally used in order to fix the free constants
$A'$ and $A$ in the interior and exterior regions respectively, thus fixing completely the
global problem in the $l=2$ sector.

We discuss finally 
the deformation of the boundary.
In \cite{Hartle1967} the analysis of the deformation needs the introduction
of a function $\xi^H(r,\theta)=\xi^H_0(r)+\xi^H_2(r) P_2(\cos\theta)$
defined in the whole interior region by imposing
$P_\pertp(R+\pertp^2\xi^H(R,\theta),\theta)=\Pb(R)$ for $R\in [0,\ro]$
(see also the discussion in \cite{Bradley_etal2007}). 
The deformation is then determined by the values $\xi^H_0(\ro)$ and $\xi^H_2(\ro)$.

Let us recall that in the present treatment the deformation is described by $\defor_0$ and $\defor_2$,
which are determined by equations (\ref{eq:defor0}) and (\ref{eq:defor2}) whenever $\Eb(\ro)\neq 0$.
In the case $\Eb(\ro)=0$ the deformation can be determined by relying on a particular gauge
in order to define extensions for both $\defor_0$ and $\defor_2$.
The correspondence of $\xi^H_0(r)$ and $\xi^H_2(r)$ as \emph{functions} defined in the interior
region with quantities in the treatment presented here rely, in fact, on the construction of
those extensions. This is discussed in Appendix \ref{sec:Rgauge}, where it is shown
how equations (\ref{eq:defor0}) and (\ref{eq:defor2}) hold in all cases,
and that 
the values $\xi^H_0(\ro)$ and $\xi^H_2(\ro)$
correspond to
\[
\xi^H_0(\ro)=-\frac{1}{2}e^{\nu(\ro)/2}\defor_0,
\qquad
\xi^H_2(\ro)=-\frac{1}{2}e^{\nu(\ro)/2}\defor_2.
\]
(The relative minus sign comes from the orientation of the normal chosen
in (\ref{normal_vector}), which goes as $-\partial_r$.)
Indeed, the former translates, via (\ref{eq:defor0}), to equation (117) in \cite{Hartle1967}, which should in fact be corrected to
$\xi^H_0(\ro)=p^H_0{}^*(\ro)\ro(\ro-2M)/M$, whose value describes
the average expansion of the shape of the star \cite{HartleThorne1968,Bradley_etal2007}.
The combination of the latter with  (\ref{eq:defor2})
enters the different definitions of the ellipticity of the star
found in the literature (see e.g. \cite{HartleThorne1968}, \cite{Bradley_etal2007}) accordingly.
In particular, it provides the expression
for the ellipticity as defined in \cite{Hartle1967} 
by $\epsilon=-\frac{3}{2\ro} \xi^H_2(\ro)$, which thus reads
\[
 \epsilon=
\frac{3(\ro-2M)}{2M}\left(\Htwo(\ro)+\frac{1}{3}\frac{\ro^3}{\ro-2M}\left(\frac{2J}{\ro^3}-\Omega_\infty\right)^2\right),
\]
in agreement with (146) in \cite{Hartle1967}.

\appendix
\section{Deformation of the surface and the $\EP$-gauge}
\label{sec:Rgauge}
We devote this Appendix to discuss 
the deformation of the surface, and at the same time, study
the relationship of the two gauges used in \cite{Hartle1967} (also in \cite{Bradley_etal2007}).

In order to describe the deformation of the surface,
motivated by the approaches taken in Newtonian theory,
it has been common in the literature to focus on the surfaces of constant energy density
or, equivalently, of constant pressure given a barotropic equation of state.
This consists after all of a choice of gauge in which
the surfaces of constant energy density (or pressure) in the interior region of the
perturbed configuration are those of constant radial coordinate.
This is described in \cite{Hartle1967} (see also \cite{Bradley_etal2007})
as a change from the original coordinate $r^H$ (the initial gauge corresponds to the $k$-gauge)
to another $R$ defined by (the inverse of) 
\begin{equation}
  \label{eq:rH_pertp}
  \{R,\theta\} \to \{r^H=r^H_\pertp(R,\theta),\theta\} 
\end{equation}
for some function $r^H_\pertp(R,\theta)$
satisfying $r^H_0(R,\theta)=R$ and
\begin{equation}
  \label{eq:PgaugeH}
  E_\pertp(r^H_\pertp(R,\theta),\theta)=\Eb(R),
\end{equation}
where $E_\pertp$ is the energy density corresponding to $g_\pertp$ (see (\ref{em-family}))
in the $k$-gauge.
The surfaces of constant energy density in the perturbed configuration, $E_\pertp$,
are then those of constant $R$, and their values correspond to
the values the pressure of the background configuration $\Eb$
take at those $R\in (0,\ro]$. 
In the present terminology that corresponds to moving to another gauge,
to which we refer to as the $\EP$-gauge.
Note that (\ref{eq:PgaugeH}) is imposed for all $\pertp$
in some neighbourhood around $0$, and therefore for all orders.
To second order $r^H_\pertp(R,\theta)$ is specified in \cite{Hartle1967} as 
\begin{equation}
  \label{eq:xiH}
  r^H_\pertp(R,\theta)=R+\pertp^2\xi^H(R,\theta)+O(\pertp^3),
\end{equation}
where for clarity we write explicitly the perturbation parameter at this point.
We do not comment yet on the existence nor uniqueness of the $\EP$-gauge.

In \cite{Hartle1967} the perturbed surface 
is then \emph{defined} as the surface of constant energy density that equals
the value of the energy density at the surface of the static configuration.
Explicitly, $\Supfamp_\pertp$ is defined to have the form
$\Supfamp_\pertp:r^H=r^H_\pertp(a,\theta)$, 
which is equivalent to $R=a$ by construction.

Let us formulate 
that condition in the present treatment.
Indicating with a ${}^{(\EP)}$ when a (gauge-dependent) quantity or object refers to the $\EP$-gauge,
the expression (\ref{eq:PgaugeH})
can be cast just as 
$$E^{(n)(\EP)}=0$$
for all orders $n\geq 1$ (note $E^{(0)(\EP)}=E$).
At each order $n$ that condition would
determine, in principle, the $\EP$-gauge at the corresponding order.
The perturbed matching hypersurface $\Supfamp_\pertp$ would then be \emph{defined}
by imposing $\Supfamp^{(\EP)}_\pertp=\Sigma_0$ pointwise
(see Section \ref{sec:overview}, and  note we are referring to the interior ($+$) region).
In other words,
the perturbed matching hypersurface is \emph{defined} by
imposing that the $\EP$-gauge is, at the same time,
a ``surface-comoving'' gauge.

Given a barotropic equation of state all the above can be stated in terms
of the pressure. The $\EP$-gauge is then also determined by 
\begin{equation}
  \label{eq:defsurface2}
   P^{(n)(\EP)}=0
\end{equation}
for all $n\geq 1$.
Since the interior pressure necessarily vanishes
at the boundary in the background configuration,
imposing that the $\EP$-gauge is also
a ``surface-comoving'' gauge implies that
the whole perturbed pressure computed in the $\EP$-gauge vanishes at the perturbed boundary.
This is the view taken in \cite{Bradley_etal2007} and many other works (see e.g. \cite{CMMR,CMMRpolitropo}).

Clearly, given a barotropic equation of state, the approach taken in terms of $E$
(say, approach ``E'')
and that in terms of $P$ (approach ``P'') lead to the same conclusion.
However, their justifications are of different nature, apart from the possible problems
of existence.

Regarding the approach ``E'', if $\Eb(\ro)\neq 0$ the fact that the perturbed energy density 
attains that value $\Eb(\ro)$  at the boundary
may, in principle and in general, seem to constitute an assumption.
Probably due to this difficulty the approach ``P'' has seemed to be preferred in many works since the vanishing of
the (perturbed) ``pressure'' on the surface is what one would expect on physical grounds.
However, that would be an erroneous statement as such, and in general,
since $P_\pertp$ is gauge dependent (see Section \ref{sec:pf-vacuum}). One should, at least, prove in which gauge that should happen.
Indeed, the matching conditions in the exact case restrict the possible jumps of the Einstein tensor across the surface. However, it remains to be shown how this fact
translates to the perturbative matching scheme in the general case.
A general consistent approach should not rely, in principle, on the use of a result (the vanishing of a ``pressure'' in a certain gauge) that has to be proven, in fact, as a consequence of the procedure.

Finally, the definition of the deformation of the star in terms of the $\EP$-gauge should control
and take care of the existence (and uniqueness, if needed) of the gauge. For instance,
in the simplest case of a constant energy density interior background
$\Eb(r)=\Eb(\ro)=const.$ the $\EP$-gauge cannot be determined using (\ref{eq:PgaugeH}), and thus,
neither the deformation. Instead, the ``P'' approach has to be used,
for which the $\EP$-gauge can be constructed. This is implicitly done in
works focused on stars of constant energy density, such as \cite{Chandra_Miller1974}.

Nevertheless, the determination of
$\Supfamp_\pertp$ using the $\EP$-gauge is well justified if
$\Eb(\ro)=0$ but $\Eb(r)\neq 0$ ($>0$ in fact) for $r\in (0,\ro)$,
since then the perturbed star (perfect-fluid region) extends up to where $E_\pertp$ vanishes,
and no further.
By the local nature of the matching, one could relax this condition
to $\Eb(\ro-\delta)\neq 0$ for all $\delta>0$ in some neighbourhood of $\ro$.
This condition (and analyticity of $\Eb(r)$) demand that there exists $n$ such that $n$-th derivative $d^n\Eb/dr^n(\ro)$
at $r=\ro$ is non-zero.
The implicit function theorem can then be applied to every differentiation of
(\ref{eq:PgaugeH}) with respect to $\pertp$ evaluated at $\pertp=0$ in order to
show that $r^H_\pertp(a,\theta)$ can be obtained order by order from (\ref{eq:PgaugeH}).
The full proof is out of the scope of this appendix and will be presented elsewhere.
When needed, we will simply assume that the $\EP$-gauge can be constructed
from $r=\ro$ inwards.

As stressed, in the present treatment no argument about the vanishing
of the pressure of the perturbed configuration $P_\pertp$ has been made, nor any specific gauge has been used.
In Sections \ref{sec:first_order} and \ref{sec:second_order_problem_matching} it has been shown how
the deformation of the boundary, described by the quantities
$Q_1$ of the first order and
$\defor_0$ and $\defor_2$ of the second order, 
are determined by $Q_1=0$ when $\Eb(\ro)\neq 0$ or 
$\Eb'(\ro)\neq 0$,
and (\ref{eq:defor0}) and (\ref{eq:defor2}) when $\Eb(\ro)\neq 0$, respectively,
and how that agrees with the results in \cite{Hartle1967}.

In what follows 
we first show explicitly that the $\EP$-gauge is indeed a ``surface gauge'' 
when $\Eb(\ro)\neq 0$, at least to second order. This shows, at the same time, that the 
usual ``vanishing of the pressure at the boundary'' in the exact case
translates in this perturbative scenario to $P^{(\EP)}_\pertp|_{\Supfamp^{(\EP)}_\pertp}=0$,
i.e. that the perturbed pressure in the $\EP$-gauge must vanish
at the perturbed surface (at least to second order).
Secondly, we use the definition of the
perturbed surface when $\Eb(\ro)=0$ by means of the $\EP$-gauge (approach ``E'') to
show that, given the $\EP$-gauge exists (and is unique), then $Q_1=0$ and equations
(\ref{eq:defor0}) and (\ref{eq:defor2}) hold even when $\Eb(\ro)= 0$.

Not to overwhelm the notation let us drop the interior
$+$ superscripts in the following when not needed.

As shown in Section \ref{sec:first_order}, at first order we have $\Eb^{(1)}=\Pb^{(1)}=0$,
and the condition $\Eb(\ro)\neq 0$ already implies $Q_1=0$. Therefore,
the family of gauges chosen for the family (\ref{gefamily})
satisfies the $\EP$-gauge condition to first order.
Since $Q_1=0$, $\Supfamp_\pertp$ coincides at first order with $\Supfamp_0$
as a set of points. The $\EP$-gauge is therefore a ``surface-comoving'' gauge
up to first order.
A \emph{hypersurface} gauge can be used to fix $\vec T_1^+=0$,
so that the perturbed $\Supfamp_\pertp$  coincides at first order with $\Supfamp_0$
pointwise, so that the $\EP$-gauge is, moreover, a ``surface'' gauge
up to first order.

Regarding the second order, let us recall that given conditions at the origin
(such that $\Ptsppz(0)$ vanishes)
$\Ptsppz(r)$ is fully determined by the $l=0$ field equations,
and $\Ptsppt(r)$ is obtained from (\ref{eq:integral_l2}),
once $\Htwo(r)$ is fully determined, in turn, by the $l=2$ field equations
and the condition at the origin and at the boundary $r=\ro$ coming from 
the ``continuity'' of the functions  $\Htwo$ and $\Ktwo$.
Now, the $\EP$-gauge is selected by fixing $\kk(r)$
and $\fr(r)$ so that $\Pppz{}^{(\EP)}(r)$ and $\Pppt{}^{(\EP)}(r)$ vanish.
From
(\ref{eq:P0gorda}) and (\ref{eq:P2gorda}) this is accomplished by imposing
\begin{equation}
\kk^{(\EP)}=-\frac{\Eb+\Pb}{r\Pb'}\Ptsppz,
\qquad
\fr^{(\EP)}=-\frac{\Eb+\Pb}{r\Pb'}\Ptsppt.
\label{eq:Rgauge_k0f2}
\end{equation}

We are ready to show that if (\ref{eq:defor0}) and (\ref{eq:defor2}) hold
then $\Qtwo^{(\EP)}=0$. This follows directly from the definitions
(\ref{eq:def_deformation}),
which in the $\EP$-gauge read
\[
\Qttwoz^{(\EP)}=\defor_0+2ae^{-\nu(\ro)/2}\kk^{(\EP)}(\ro),\qquad
\Qttwot^{(\EP)}=\defor_2+2ae^{-\nu(\ro)/2}\fr^{(\EP)}(\ro).
\]
Equations (\ref{eq:defor0}) and (\ref{eq:defor2}) together with (\ref{eq:Rgauge_k0f2})
evaluated on $r=\ro$ readily imply $\Qttwoz^{(\EP)}=\Qttwot^{(\EP)}=0$.
Finally, since we have chosen $\vec T_1=0$ at first order, then
$\Qtwoz^{(\EP)}=\Qtwot^{(\EP)}=0$ as follow from the definitions (\ref{eq:def_Qt2}).
It only remains, again, to choose a convenient \emph{hypersurface} gauge
to second order to fix $\vec T_2^+=0$ so that the perturbed
$\Supfamp_\pertp$ coincides with $\Supfamp_0$ at second order, not only
as a set of points, but pointwise.
We have thus shown that \emph{the $\EP$-gauge is indeed a ``surface gauge''  
whenever $\Eb(\ro)\neq 0$, at least to second order}, as expected.

Let us consider now the case
$\Eb(\ro)=0$ under the conditions that ensure the existence and
construction of the $\EP$-gauge.
The matching hypersurface $\Supfamp_\pertp$
is then determined 
by the coincidence of $\Supfamp^{(\EP)}_\pertp$ and $\Sigma_0$
pointwise
 (in $\mmm_0^+$, mind the $+$ superscript).
This condition is equivalent, up to second order, to $Q_1^{+(\EP)}=\Qtwo^{+(\EP)}=0$
together with a hypersurface gauge choice such that $\vec T_1^+=\vec T_2^+=0$ at each order.
At first order we thus have the required result by construction.
At second order,
the equations defining $\defor_{0/2}$ (\ref{eq:def_deformation}) in the interior
read then
\[
\defor_0=-2ae^{-\nu(\ro)/2}\kk^{(\EP)}(\ro),\qquad
\defor_2=-2ae^{-\nu(\ro)/2}\fr^{(\EP)}(\ro),
\]
which combined with (\ref{eq:Rgauge_k0f2}),
yield (\ref{eq:defor0}) and (\ref{eq:defor2}).

We must finally address the issue of how
$\defor_{0/2}$, given by
(\ref{eq:defor0}) and (\ref{eq:defor2}), describe the deformation
of the surface.
The key is to show how the deformation quantities $\defor_{0/2}$,
defined on $\Supfamp_0$,
can be extended to the interior region and how that relates to the change from the $k$-gauge to the $\EP$-gauge.
We start by defining that change in terms of $\vec V_2$.
Let us, for simplicity, set $\vec s_1=0$ so that $\vec s_2=\vec V_2$.
Including $\vec s_1=Ct\partial_\varphi$ does not add anything relevant to the analysis.
Recall that the $k$-gauge is defined by $\kk^{(k)}=0$ and $\fr^{(k)}=0$.
Given that the second order change $\vec V_2=2S(r,\theta)\partial_r$ induces (\ref{eq:K2_s2}) (with $C=0$),
it is immediate to check (recall the freedom in defining $f(r,\theta$)) 
that the change from the $k$-gauge to the $\EP$-gauge is accomplished by setting
\begin{equation}
\vec V_2=2r \left(\kk^{(\EP)}+\fr^{(\EP)} P_2(\cos\theta)\right) \partial_r
=-2\frac{\Eb+\Pb}{\Pb'}\left(\Ptsppz+\Ptsppt P_2(\cos(\theta)\right)\partial_r,
\label{eq:V2Rgauge}
\end{equation}
where the second equality follows from (\ref{eq:Rgauge_k0f2}).
Note that the relation $\kkt^{(k)}=\kkt^{(\EP)}-\fr^{(\EP)}$ holds.

On the other hand, given the definition of the second order gauge vectors in (\ref{eq:s_gauges}),
the second order gauge $\vec V_2=2S(r,\theta)\partial_r$ with $\vec s_1=0$
corresponds to a diffeomorphism $\Omega_\pertp:\mmm_0\to\mmm_0 $ of the form $(s,\theta)\to (\mathcal{R}_\pertp(s,\theta),\theta)$
for $s\in[0,\ro]$ defined by $\mathcal{R}_\pertp(s,\theta)=s+\pertp^2 S(s,\theta)$.
Given (\ref{eq:V2Rgauge}), we thus have 
\begin{equation}
  \label{eq:diff_gauge}
  \mathcal{R}_\pertp(s,\theta)=s- \pertp^2\frac{\Eb(s)+\Pb(s)}{\Pb'(s)}
\left(\Ptsppz(s)+\Ptsppt(s) P_2(\cos\theta)\right).
\end{equation}
Let us recall again (see Section \ref{sec:conclusion}) that the
coordinate $R$ used in \cite{Hartle1967} ranges from
$0$ to $a$, and therefore (\ref{eq:diff_gauge}) can be compared with the expression (\ref{eq:xiH})
in the form $r^H_\pertp(s,\theta)=s+\pertp^2\xi^H(s,\theta)+O(\pertp^3)$ to obtain
\[
\xi^H=-\frac{\Eb+\Pb}{\Pb'}
\left(\Ptsppz+\Ptsppt P_2(\cos\theta)\right).
\]
Now, this is in agreement with 
$\xi^H=\xi^H_0+\xi^H_2 P_2(\cos\theta)$ for
$\xi^H_{0/2}=-\frac{\Eb+\Pb}{\Pb'}p^H_{0/2}{}^*$, as follows from (90) and (91) in \cite{Hartle1967}
and the correpondences $p^H_{0/2}{}^*(s)=\Ptspp_{0/2}(s)$ found in Section \ref{sec:conclusion}.

Expression (\ref{eq:diff_gauge}) suggests the construction of two functions
in the interior
\begin{equation}
  \label{eq:Rgauge_xis}
\xi_{0/2}\defi 2\frac{\Eb+\Pb}{\Pb'} e^{-\nu/2} \Ptspp_{0/2}.
\end{equation}
These, evaluated at $r=\ro$, and given that (\ref{eq:defor0}) and (\ref{eq:defor2}) hold,
lead to
\[
\xi_{0/2}(\ro)=\defor_{0/2}. 
\]
The functions $\xi_{0/2}$ (\ref{eq:Rgauge_xis}) are therefore extensions of $\defor_{0/2}$, as \emph{defined}
in (\ref{eq:defor0}) and (\ref{eq:defor2}), to all the interior region, and are `radial'-gauge independent by construction.
The information of the deformation of the star in the $k$-gauge is therefore encoded in
the functions $\xi_{0/2}$, whereas  in the $\EP$-gauge that information lies in the functions $\kk^{(\EP)}$ and $\fr^{(\EP)}$.

Using the correspondence $\xi_{0/2}(s)=-2e^{-\nu/2}\xi^H_{0/2}(s)$, so that $\defor_{0/2}=-2e^{-\nu/2}\xi^H_{0/2}(a)$,
the analysis of the deformation of the star in terms of $\defor_{0}$ and $\defor_{2}$
follows then from the discussions in \cite{Hartle1967} (see also \cite{Bradley_etal2007}).
Note that the minus sign in the correspondence comes from the choice
of the normals as defined in (\ref{normal_vector}), which point towards the origin.

\section{Proofs of Propositions \ref{teo:first_order_matching}
and \ref{teo:second_order_matching}}
\label{sec:proofs}
For the sake of completeness we include in this Appendix the explicit expressions needed
in order to use Theorem 1 from \cite{Mars2005},  as obtained from Propositions
2 and 3 from that reference, particularized for a timelike hypersurface $\Supfamp_0$.
Let us start by decomposing $\fpt$ in its normal and tangent parts with respect to $\vec n$
as
\[
\Kper_{\alpha \beta} = \Kpernornor n_{\alpha} n_{\beta} + n_{\alpha} \Kpernortan_{\beta} +
n_{\beta} \Kpernortan_{\alpha} + \Kpertan_{\alpha\beta},
\]
where the vector $\Kpernortan_{\alpha}$ and symmetric tensor $\Kpertan_{\alpha\beta}$ denote the projected
components of $\Kper_{\alpha \beta}$ on $\Sigma_0$, that is $\Kpernortan_{\alpha}n^\alpha=0$ and $\Kpertan_{\alpha\beta}n^\alpha=0$.

The first and second fundamental forms to first order, $\qper$ and $\kappaper$, are given by the expressions
\cite{Mars2005}
\begin{eqnarray}
\qper_{ij} & \! \! = \! \! &  
\lie_{\vec{\Tone} } h_{ij}  + 2 \Qone \kfamp_{ij} + \Kper_{\alpha\beta} e^{\alpha}_i e^{\beta}_j,\label{h1ij}
\\
\kappaper_{ij} & \! \! = \! \! &  
\lie_{\vec{\Tone} } \kfamp_{ij} 
- D_{i} D_{j} \Qone 
+ \Qone \left ( - n^{\mu} n^{\nu} R_{\alpha\mu\beta\nu}
e^{\alpha}_i e^{\beta}_j + \kfamp_{il} \kfamp^{\,\,l}_{j} \right )
+ \frac{1}{2}  \Kpernornor 
\kfamp_{ij} - n_{\mu} \Sper^{\mu}_{\alpha\beta}e^{\alpha}_i e^{\beta}_j, \nonumber
\end{eqnarray}
where $D_i$ is the three dimensional covariant derivative of $(\Sigma_0,\hfamp)$  and
\[
\Sper^{\alpha}_{\beta\gamma}  \equiv \frac{1}{2} \left  ( \nabla_{\beta}
\Kper^{\alpha}_{\,\gamma}  +  \nabla_{\gamma}  \Kper^{\alpha}_{\,\,\beta}  -
\nabla^{\alpha}  \Kper_{\beta\gamma}  \right  ).
\]
The first and second fundamental forms to second order, $\qperper$ and $\kappaperper$, are given by the expressions \cite{Mars2005}
\begin{eqnarray}
&&\qperper_{ij} = \lie_{\vec{\Ttwo}} h_{ij} + 2 \Qtwo \kappa_{ij} + \Kperper_{\alpha\beta} e^{\alpha}_i e^{\beta}_j
+ 2 \lie_{\vec{\Tone}} \qper_{ij} - \lie_{\vec{\Tone}} \lie_{\vec{\Tone}} h_{ij}  \nonumber \\
&&\quad+ \lie_{2 \Qone \vec{\Kpernortan} - 2 \Qone \kappa (\vec{\Tone} )   -  D_{\vec{\Tone}} \vec{\Tone} } h_{ij} 
+ 2 \left ( \Tone^{l}  \Tone^{s} \kappa_{ls}
- 2 \vec{\Tone} ( \Qone )  + 2 \Qone \Kpernornor \right )
\kappa_{ij}  \nonumber \\
&&\quad+ 2 \Qone^2 \left ( -n^{\mu} n^{\nu} R_{\alpha\mu\beta\nu} e^{\alpha}_i
e^{\beta}_j + \kappa_{il} \kappa^{l}_{j} \right ) + 2  D_{i} \Qone D_{j} \Qone
- 4 \Qone n_{\mu} \Sper^{\mu}_{\alpha\beta} e^{\alpha}_i e^{\beta}_j
,\nonumber\\ 
\nonumber \\
&&\kappaperper_{ij} =
\lie_{\vec{\Ttwo}} \kappa_{ij} -  D_{i} D_{j} \Qtwo 
- \Qtwo n^{\mu} n^{\nu} R_{\alpha\mu\beta\nu} e^{\alpha}_i e^{\beta}_j 
+ \Qtwo \kappa_{il} \kappa^{l}_{k} 
- n_{\mu} \Sperper^{\mu}_{\alpha\beta} e^{\alpha}_i e^{\beta}_j 
 \nonumber \\
&&\quad+ 2 \lie_{\vec{\Tone}} \kappaper_{ij}
+ \kappa_{ij} \left ( \frac{1}{2} \Kperpernornor - \frac{1}{4} \Kpernornor^2 -  
\left ( \tau_{l} + D_{l}\Qone \right )
\left ( \tau^{l} + D^{l}\Qone \right ) 
+ 2 \Qone n_{\mu} n^{\rho} n^{\delta}\Sper^{\mu}_{\rho\delta} \right )
\nonumber \\
&&\quad+  \left ( \Kpernornor n_{\mu} + 2 \tau_{\mu} + 2 D_{\mu} \Qone \right ) 
\Sper^{\mu}_{\alpha\beta} e^{\alpha}_i e^{\beta}_j
- 2 \Qone n_{\mu} n^{\nu} ( \nabla_{\nu} \Sper^{\mu}_{\alpha\beta} ) e^{\alpha}_i e^{\beta}_j
- 2 n_{\mu} n^{\nu} \Sper^{\mu}_{\alpha \nu} e^{\alpha}_i D_j \Qone \nonumber \\
&&\quad- 2 n_{\mu} n^{\nu} \Sper^{\mu}_{\alpha \nu} e^{\alpha}_j D_i \Qone 
- 2 \Qone n_{\mu} \Sper^{\mu}_{\alpha \beta} e^{\alpha}_i e^{\beta}_l \kappa^{l}_j
- 2 \Qone n_{\mu} \Sper^{\mu}_{\alpha \beta} e^{\alpha}_j e^{\beta}_l \kappa^{l}_i
\nonumber \\ 
&&\quad+ \lie_{\grad (\vec{\Tone} (\Qone)) 
- \frac{1}{2} \grad (\Tone^{l} \Tone^{m} \kappa_{lm} )
- \frac{1}{2} \Kpernornor \grad (\Qone) + 2 \Qone \kappa ( \grad \Qone )} \,  h_{ij}
\nonumber \\ 
&&\quad+  \left ( 2 \vec{\Tone} (\Qone) - \Tone^{l} \Tone^{m} \kappa_{lm} -
\Qone \Kpernornor \right )\left ( n^{\mu} n^{\nu} R_{\alpha\mu\beta\nu}
e^{\alpha}_i e^{\beta}_j - \kappa_{il} \kappa^{l}_{j}  \right ) 
+ \frac{1}{2} \left ( D_{i} \Qone D_j \Kpernornor + \right . \nonumber \\
&&\quad\left. + D_{j} \Qone D_i \Kpernornor \right ) 
-  \lie_{\vec{\Tone}} \lie_{\vec{\Tone}} \kappa_{ij} 
-  \lie_{2 \Qone \kappa (\vec{\Tone} ) +  D_{\vec{\Tone}} \vec{\Tone} } \, \kappa_{ij} 
- 2 \Qone \lie_{\grad (\Qone) } \kappa_{ij}
\nonumber \\
&&\quad- \Qone^2 \left ( 
n^{\mu} n^{\nu} n^{\delta} ( \nabla_{\delta} R_{\alpha\mu\beta\nu} ) e^{\alpha}_i e^{\beta}_j
+ 2 n^{\mu} n^{\nu} R_{\delta\mu\alpha\nu} e^{\delta}_l e^{\alpha}_j \kappa^{l}_i
+ 2 n^{\mu} n^{\nu} R_{\delta\mu\alpha\nu} e^{\delta}_l e^{\alpha}_i \kappa^{l}_j \right ),\nonumber
\end{eqnarray}
where $\Kperpernornor=\spt_{\alpha\beta}n^\alpha n^\beta$,
\[
\Sperper^{\alpha}_{\beta\gamma}  \equiv \frac{1}{2} \left  ( \nabla_{\beta}
\Kperper^{\alpha}_{\,\gamma}  +  \nabla_{\gamma}  \Kperper^{\alpha}_{\,\,\beta}  -
\nabla^{\alpha}  \Kperper_{\beta\gamma}  \right  ),
\]
and for any tangent vector $\vec V$, $\kfamp(\vec V)$ is the vector $\kfamp^i{}_jV^j$.

\subsection{Proof of Proposition \ref{teo:first_order_matching}}
Theorem 1 in \cite{Mars2005} states that the first order matching conditions are satisfied if there exist
two scalars $Q_1^\pm$ and two vectors $\vec T_1^\pm$ on $\Sigma_0$
such that the system of equations given by $\qper^+_{ij}=\qper^-_{ij}$ and $\kappaper^+_{ij}=\kappaper^-_{ij}$
admits a solution.
We start by calculating $\qper$ and  $\kappaper$ through expressions (\ref{h1ij}).
Let us recall these are objects defined on $\Sigma_0$, which is non-degenerate.
The ingredients needed are the
background embeddings (\ref{sigma0+}), (\ref{sigma0-}), with tangent basis (\ref{eq:tangents})
and unit normals 
(\ref{normal_vector}),
plus the first and second fundamental forms of $\Sigma_0$  (\ref{h0ij}),
together with the first order perturbation tensors $\fpt^\pm$ (\ref{fopert_tensor}) restricted to $\Sigma_0$ at each side. The functions $\Qone(\tau,\vartheta)$
and vectors $\vec{T}^\pm_1={T_1^\tau}^\pm (\tau, \vartheta) \partial_\tau + {T_1^\phi}^\pm (\tau, \vartheta) \partial_\phi
+{T_1^\vartheta}^\pm (\tau, \vartheta) \partial_\vartheta$ on $\Sigma_0$ at each side are left as unknowns.
The explicit expressions of $\qper^\pm$ and $\kappaper^\pm$ read
\begin{eqnarray*}
\qper{}^\pm_{ij}dx^idx^j&=&e^{\nu(\ro)}\left(-2T_1^\pm{}^\tau+\nu'(\ro)e^{-\frac{\lambda(\ro)}{2}}\Qone^\pm\right)d\tau^2
+2\left(-e^{\nu(\ro)}T_1^\pm{}^\tau_{,\vartheta}+\ro^2T_1^\pm{}^\vartheta_{,\tau}\right) d\tau d\vartheta\\
&&+2\ro^2\left(T_1^\pm{}^\phi_{,\tau}-\omega^\pm(\ro,\vartheta)\right)\sin^2\vartheta d\tau d\phi\\
&&+2\ro(\ro T_1^\pm{}^\vartheta_{,\vartheta} -e^{-\frac{\lambda(\ro)}{2}}\Qone^\pm)d\vartheta^2
+2\ro^2T_1^\pm{}^\phi_{,\vartheta}\sin^2\vartheta d\vartheta d\phi\\
&&+2\ro\left( \ro T_1^\pm{}^\vartheta\cos\vartheta- e^{-\frac{\lambda(\ro)}{2}}\Qone^\pm\sin\vartheta\right)\sin\vartheta d\phi^2,
\end{eqnarray*}
\begin{eqnarray*}
&&\kappaper{}^\pm_{ij}dx^idx^j=\\
&&\left\{-\Qone^\pm{}_{,\tau\tau}+e^{-\frac{\lambda(\ro)}{2}}e^{\nu(\ro)}
\left(T_1^\pm{}^\tau_{,\tau}+e^{-\frac{\lambda(\ro)}{2}}\frac{\Qone^\pm}{4}\left(\lambda'_\pm(\ro)\nu'(\ro)
-2\nu''_\pm(\ro)-2\nu'^2(\ro)\right)\right)\right\}d\tau^2\\
&&-2\left\{\Qone^\pm{}_{,\tau\vartheta}+e^{-\frac{\lambda(\ro)}{2}}
\left(\ro T_1^\pm{}^\vartheta_{,\tau}-\frac{1}{2}e^{\nu(\ro)}\nu'(\ro)T_1^\pm{}^\tau_{,\vartheta}\right)\right\}d\tau d\vartheta\\
&&+2\ro e^{-\frac{\lambda(\ro)}{2}}\left(-T_1^\pm{}^\phi_{,\tau}
+\frac{1}{2}\ro \omega'^\pm(\ro,\vartheta)+\omega^\pm(\ro,\vartheta)\right)\sin^2\vartheta d\tau d\phi\\
&&-\left\{\Qone^\pm{}_{,\vartheta\vartheta}+2\ro T_1^\pm{}^\vartheta_{,\vartheta}
+e^{-\lambda(\ro)}\Qone^\pm \left(\frac{1}{2}\ro\lambda'_\pm(\ro)-1\right) \right\}d\vartheta^2\\
&&-2\ro e^{-\frac{\lambda(\ro)}{2}}T_1^\pm{}^\phi_{,\vartheta}\sin^2\vartheta  d\vartheta d\phi\\
&&-\left\{\left(\Qone^\pm{}_{,\vartheta}+2\ro e^{-\frac{\lambda(\ro)}{2}} T_1^\pm{}^\vartheta \right)
\cos\vartheta
+e^{-\lambda(\ro)}\Qone^\pm\left(\frac{1}{2}\ro\lambda'_\pm(\ro)-1\right)\sin\vartheta\right\}\sin\vartheta d\phi^2,
\end{eqnarray*}
where the background matching conditions (\ref{background_matching}) have been used
to set $\nu_\pm(\ro)=\nu(\ro)$, $\nu'_\pm(\ro)=\nu'(\ro)$ and $\lambda_\pm(\ro)=\lambda(\ro)$.

The ordered procedure used in order to obtain and integrate the difference functions is
the following.
First, from $[\qper_{\vartheta\phi}]=0$ we obtain $[\Tone^\phi]_{,\vartheta}=0$.
On the other hand, the derivative
$[\qper_{\tau\phi}]_{,\tau}=0$
yields $[\Tone^\phi]_{,\tau\tau}=0$, and therefore
$[\Tone^\phi]=b_1\tau+C_2$
for arbitrary constants $b_1$ and $C_2$.
As a result, $[\qper_{\tau\phi}]=0$ reads
$
[\omega]=b_1.
$

Now,  equation $[\qper_{\vartheta\vartheta}]\sin^2\vartheta-[\qper_{\phi\phi}]=0$ yields
$[\Tone^\vartheta]\cos\vartheta - [\Tone^\vartheta]_{,\vartheta}\sin\vartheta=0$,
which is integrated into $[\Tone^\vartheta]=F(\tau) \sin\vartheta$ for some function $F(\tau).$
Equation $[\qper_{\vartheta\vartheta}]=0$  now reads
$[\Qone]=e^{\lambda(\ro)/2}\ro F\cos\vartheta$.
On the other hand, the compatibility condition to integrate $[\Tone^\tau]$ is given by
$2[\qper_{\tau\vartheta}]_{,\tau}-[\qper_{\tau\tau}]_{,\vartheta}=0$,
which yields $\ddot F=-F\nu'(\ro)e^{\nu(\ro)}/2\ro$, and thence
$[\Tone^\tau]=C_1- e^{-\nu(\ro)}\ro^2\dot F\cos\vartheta$ for some arbitrary constant $C_1$.
We have so far exhausted the conditions $[\qper_{ij}]=0$.

Given the above conditions, equation $[\kappaper_{\tau\phi}]=0$ is now equivalent to
$
[\omega']=0.
$
The conditions on the metric perturbations have thus been obtained.

Consider the equation $[\kappaper_{\tau\vartheta}]=0$, which now reads
$\dot F\ro\sin\ro(2e^{\lambda(\ro)}-2+\ro\nu'(\ro))=0$.
If $2e^{\lambda(\ro)}-2+\ro\nu'(\ro)\neq 0$ we then have
$\dot F=0$, which due to its previous equation can only be satisfied in the trivial case $F=0$.
From the above, in particular, $[\Qone]=0$.
Then, equations $[\kappaper_{\phi\phi}]=0$ and $[\kappaper_{\vartheta\vartheta}]=0$
just provide $\Qone[\lambda']=0$, from which $[\kappaper_{\tau\tau}]=0$
thus reads $\Qone[\nu'']=0$.

The appearance of the constants
$C_1$ and $C_2$  
is a consequence of the isometries present in the background
configuration, and cannot be determined \cite{Mars2007}. Nevertheless, they can be safely
absorbed by using a isomorphic \emph{spacetime} gauge at one (any) side,
say $\vec s_1^+=C_1\partial_{\tp}+C_2\partial_{\phip}$, which,
by (\ref{eq:Z_gauges}) leads to $\vec T^+_1\to \vec T^+_1-\vec s_1^+$ and
obviously leaves the metric perturbation tensor
$\fpt^+$ unchanged. We can thus set $C_1=C_2=0$ without loss of generality.
\fin

\subsection{Proof of Proposition \ref{teo:second_order_matching}}
The procedure is analogous to that of the previous proof.
We first consider the case $\left[\lambda'\right]\neq 0$ or
$\left[\nu''\right]\neq 0$, so that $\Qone=0$ necessarily.
The explicit expression of $\qperper^\pm$ reads
\begin{eqnarray*}
  &&\qperper{}^\pm_{ij}dx^idx^j=
  \left\{-2e^{\nu(\ro)}\left(T_2^\pm{}^\tau_{,\tau}
      +(T_1^\tau{}_{,\tau})^2\right)+2\ro^2\left(T_1^\pm{}^\phi_{,\tau}-\omega^\pm(\ro,\vartheta)\right)^2\sin^2\vartheta\right.\\
  &&\left.+2\ro^2(T_1^\vartheta{}_{,\tau})^2-4 e^{\nu(\ro)} h(\ro,\vartheta)+e^{-\frac{\lambda(\ro)}{2}}e^{\nu(\ro)}
    \nu'(\ro)\Qt2 \right\}d\tau^2\\
  &&+2\left\{2\ro^2 T_1^\pm{}^\phi_{,\tau}T_1^\pm{}^\phi \cos\vartheta\sin\vartheta
    -e^{\nu(\ro)}T_2^\pm{}^\tau_{,\vartheta}+\ro^2 T_2^\pm{}^\vartheta_{,\tau}
    +2\ro^2 T_1^\vartheta{}_{,\tau}T_1^\vartheta{}_{,\vartheta}\right.\\
  &&\left.+2\ro^2T_1^\pm{}^\phi_{,\vartheta}\left(T_1^\pm{}^\phi_{,\tau}-\omega^\pm(\ro,\vartheta)\right)\sin^2\vartheta
    -2e^{\nu(\ro)}T_1^\tau{}_{,\tau}T_1^\tau{}_{,\vartheta} \right\} d\tau d\vartheta\\
  &&+2\ro^2\left\{2\left(T_1^\pm{}^\phi_{,\tau}T_1^\vartheta -T_1^\pm{}^\phi T_1^\vartheta{}_{,\tau}
      -2\omega^\pm(\ro,\vartheta)T_1^\vartheta\right)\cos\vartheta\right.\\
  &&\left.+\left(T_2^\pm{}^\phi_{,\tau} -2T_1^\tau{}_{,\tau}\omega^\pm(\ro,\vartheta) -2T_1^\vartheta\omega^\pm_{,\vartheta}(\ro,\vartheta)\right)\sin\vartheta\right\}\sin\vartheta d\tau d\phi\\
  &&+2\left\{\ro^2\left(T_1^\pm{}^\phi\cos\vartheta 
      +T_1^\pm{}^\phi_{,\vartheta}\sin\vartheta  \right)^2
    -\ro^2\sin^2\vartheta(T_1^\pm{}^\phi)^2+\ro^2(T_1^\vartheta{}_{,\vartheta})^2+\ro^2 T_2^\pm{}^\vartheta_{,\vartheta}\right.\\
  &&\left. -e^{\nu(\ro)}(T_1^\tau{}_{,\vartheta})^2
    +2\ro^2 k(\ro,\vartheta)-e^{-\frac{\lambda(\ro)}{2}}\ro\Qt2^\pm \right\} d\vartheta^2\\
  &&+2\ro^2\left\{2 T_1^\vartheta T_1^\pm{}^\phi
    +\left(T_2^\pm{}^\phi_{,\vartheta} 
      -2T_1^\tau{}_{,\vartheta}\omega^\pm(\ro,\vartheta)\right)\sin^2\vartheta\right.\\
  &&  \left.+2\left(T_1^\pm{}^\phi_{,\vartheta} T_1^\vartheta - T_1^\pm{}^\phi T_1^\vartheta{}_{,\vartheta}\right)\cos\vartheta\sin\vartheta
  \right\} d\phi d\vartheta\\
  && + 2\left\{\ro^2(T_1^\pm{}^\phi)^2\cos^2\vartheta\sin^2\vartheta
    +\ro^2(T_1^\vartheta)^2(1-2\sin^2\vartheta)+\ro^2T_2^\pm{}^\vartheta\cos\vartheta\sin\vartheta \right.\\
  &&\left. +\left(2\ro^2 k(\ro,\vartheta)-e^{-\frac{\lambda(\ro)}{2}}\ro\Qt2^\pm\right)\sin^2\vartheta \right\} d\phi^2,
\end{eqnarray*}
where we have avoided the use of $\pm$ for quantities which already coincide at both sides
and we have used $\Qt2^\pm$, as defined in (\ref{eq:def_Qt2}), instead of the original $Q_2^\pm$.

From equations $[\qperper_{\tau\phi}]=0$ and $[\qperper_{\vartheta\phi}]=0$ we obtain
expressions for $[T_2^\phi]_{,\tau}$ and $[T_2^\phi]_{,\vartheta}$ respectively.
The integrability conditions are found to be automatically satisfied. The integration leads to
\begin{equation}
[T_2^\phi]= 2b_1(T_1^\tau + \tau T_1^\vartheta \cot\vartheta)+ D_2
\label{app:z2p}
\end{equation}
for some constant $D_2$.
Likewise, from $[\qperper_{\tau\vartheta}]=0$ and
$[\qperper_{\vartheta\vartheta}]\sin^2\vartheta -[\qperper_{\phi\phi}]=0$ we obtain, respectively,
$[T_2^\vartheta]_{,\tau}$ and $[T_2^\vartheta]_{,\vartheta}$. However, this time the integrability
condition provides a second order PDE for $[T_2^\tau]$, with derivatives on $\vartheta$ only, which
is integrated to yield
\begin{equation}
[T_2^\tau]=-\ro^2 \dot F(\tau) e^{-\nu(\ro)}\cos\vartheta+ G(\tau)
\label{app:z2t}
\end{equation}
for some functions $F(\tau)$, conveniently arranged, and $G(\tau)$.
$[T_2^\vartheta]$ can now be integrated in the form
\begin{equation}
[T_2^\vartheta]=\left(b_1\tau\cos\vartheta(b_1\tau-2T_1^+{}^\phi)
+F(\tau)+C_3\right)\sin\vartheta,
\label{app:z2th}
\end{equation}
for some constant $C_3$.

Now, $[\qperper_{\vartheta\vartheta}]=0$ provides an equation for $[\Qt2]$, explicitly
\begin{equation}
[\Qt2]=\ro e^{\lambda(\ro)/2}\{(2\difk+(F(\tau)+C_3)\cos\vartheta\}.
\label{app:difq2}
\end{equation}
The remaining equation from the equality of the second order first fundamental forms
is $[\qperper_{\tau\tau}]=0$. From its second
derivative $[\qperper_{\tau\tau}]_{,\tau\vartheta}=0$ we first obtain a
third order differential equation for $F(\tau)$ which can be integrated once
in order to obtain
\begin{equation}
\ddot F=e^{\nu(\ro)}\nu'(\ro)(-F+H_1-C_3)/2\ro,
\label{ddotF}
\end{equation}
where the constant of
integration $H_1$ has been conveniently arranged.
Using this relation back into the equation $[\qperper_{\tau\tau}]_{,\tau}=0$ we
obtain $\ddot G=0$, and therefore $G(\tau)=-H_0\tau+ D_1$ for some constants
$H_0$ and $D_1$. 
Finally, $[\qperper_{\tau\tau}]=0$ provides a relation
between $[h]$ and $[k]$, namely
$\difh=\frac{1}{2}H_0+\frac{1}{4}\ro \nu'(a)\left\{2\difk+H_1\cos\vartheta\right\}$.

We have to impose now the  equations for the perturbed second fundamental form,
$[\kappaperper_{ij}]=0$. 
The steps taken to solve the system of equations
are given with enough detail in what follows so that the proof can be
reproduced directly. Not to overwhelm the text we thus prefer not to include
the explicit expressions of $\kappaperper_{ij}^\pm$ here.

Firstly, given that $[\omega']=0$, the equations $[\kappaperper_{\vartheta\phi}]=0$ and
$[\kappaperper_{\tau\phi}]=0$ are automatically satisfied.
We start with the equation $[\kappaperper_{\tau\vartheta}]=0$,
which yields
$
\dot F\left(2-2e^{\lambda(\ro)}-\ro\nu'(\ro)\right)=0.
$
Since  $2-2e^{\lambda(\ro)}-\ro\nu'(\ro)\neq 0$ by assumption, we need
$\dot F=0$, and therefore, from
(\ref{ddotF}) we obtain $F+C_3=H_1$, which substituted on
the above expressions for $[T_2^\tau]$, $[T_2^\vartheta]$ and $[\Qt2]$
leads to
\begin{eqnarray}
&&[T_2^\tau]=-H_0\tau+ D_1,\label{app:z2tf}\\
&&[T_2^\vartheta]=\left(b_1\tau\cos\vartheta(b_1\tau-2T_1^+{}^\phi)
+H_1\right)\sin\vartheta,\label{app:z2thf}\\
&&[\Qt2]=\ro e^{\lambda(\ro)/2}\{(2\difk+H_1\cos\vartheta\}.\label{app:q2f}
\end{eqnarray}

On the other hand,
the combination of equations
$[\kappaperper_{\vartheta\vartheta}]\sin^2\vartheta
-[\kappaperper_{\phi\phi}]=0$, which yields a second order PDE involving $[k]-[f]$,
with derivatives on $\vartheta$ only,
is integrated to obtain $[k]=c_1(\tau)\cos\vartheta+c_2(\tau)+[f]$ for some functions
$c_1(\tau)$ and $c_2(\tau)$. However, since $\difk_{,\tau}=\difg_{,\tau}=0$, we readily have that
$c_1(\tau)=c_1$ and $c_2(\tau)=c_2$ must be constants.
Now, the equation $[\kappaperper_{\vartheta\vartheta}]=0$ provides
an expression for $\difm$, which left in terms of $\difk$, in particular, can be arranged as equation (\ref{eq:mc:m}).

The only remaining equation is given by $[\kappaperper_{\tau\tau}]=0$.
Using (\ref{eq:mc:m}) to substitute $\difm$ in $[\kappaperper_{\tau\tau}]=0$ we obtain
a relation between $\difhp$, $\difkp$ and $\difk$ (and $\Qt2^+$). That relation is given
explicitly by equation (\ref{eq:mc:hp}).

Furthermore, from the above expression for $[\Qt2]$ we clearly also obtain that
the difference
$[\Qt2]$ cannot depend on $\tau$. For the same reason, using the above equations 
for $\difm$ and $\difhp$, and since either $\left[\lambda'\right]\neq 0$
or $\left[\nu''\right]\neq 0$, then $\Qt2^+$ (and thus neither $\Qt2^-$)
cannot depend on $\tau$.

In the case $\left[\lambda'\right]=\left[\nu''\right]= 0$
we can have, in principle, a non vanishing $Q_1(\tau,\vartheta)$.
The appearance of $Q_1(\tau,\vartheta)$ in the expressions
for $\qperper_{ij}$ does not change 
the procedure to integrate the differences.
For that reason, and due to their length,
we avoid including the explicit expressions of $\qperper_{ij}$ with
$Q_1(\tau,\vartheta)\neq 0$.  
Equations $[\qperper_{\tau\phi}]=0$ and $[\qperper_{\vartheta\phi}]=0$ provide
expressions for $[T_2^\phi]_{,\tau}$ and $[T_2^\phi]_{,\vartheta}$,
the integrability conditions are automatically satisfied, and the integration leads to
the expression
$$\left[T_2^\phi\right]= 2b_1(T_1^\tau + \tau T_1^\vartheta \cot\vartheta)
-\frac{2}{\ro}e^{-\lambda(\ro)/2}b_1\tau Q_1^++D_2,
$$
for some constant $D_2$.
Now, the remaining equations in the set $[\qperper_{ij}]=0$ show no terms
involving $Q_1$. Therefore we obtain the same set of equations
(\ref{app:z2t}), (\ref{app:z2th}), (\ref{app:difq2}), (\ref{ddotF}),
$G(\tau)=-H_0\tau+ D_1$ for some constants
$H_0$ and $C_1$, and $\difh$ is given by $\difh=\frac{1}{2}H_0+\frac{1}{4}\ro \nu'(a)\left\{2\difk+H_1\cos\vartheta\right\}$.
The equation $[\kappaperper_{\lambda\vartheta}]=0$ reads the same
as in the $Q_1=0$ case, and therefore
the condition $\dot F(\tau)=0$, assuming that $2-2e^{\lambda(\ro)}-\ro\nu'(\ro)\neq 0$,
is just recovered. That again leads to $F+C_3=H_1$.
As a result $[T_2^\tau]$, $[T_2^\vartheta]$ and $[\Qt2]$, are also given by
(\ref{app:z2tf}), (\ref{app:z2thf}) and (\ref{app:q2f}).

Likewise, the combination of equations
$[\kappaperper_{\vartheta\vartheta}]\sin^2\vartheta
-[\kappaperper_{\phi\phi}]=0$ does not depend on $Q_1$ either,
and therefore $[k]=c_1\cos\vartheta+c_2+[f]$ all the same, for some constants
$c_1$ and $c_2$. However, the equation $[\kappaperper_{\vartheta\vartheta}]=0$
does contain a term involving $Q_1$. The expression for
$\difm$ in this case is given by
\begin{eqnarray}
&&\difm=\ro \difkp-\frac{1}{4}e^{-\lambda(\ro)}\left[\lambda''\right](Q_1)^2
  +\frac{1}{4}\left(\ro\lambda'(\ro)+2\right)\left\{2\difk + H_1\cos\vartheta\right\}\nonumber\\
&&\qquad-\frac{1}{2}(H_1+2c_1)e^{\lambda(\ro)}\cos\vartheta,
\label{app:difm}
\end{eqnarray}
which used in $[\kappaperper_{\tau\tau}]=0$ provides the following expression of $\difhp$
\begin{eqnarray}
&&\difhp=\frac{1}{2}\ro \nu'(\ro) \difkp-\frac{1}{4}e^{-\lambda(\ro)}\left[\nu'''\right](Q_1)^2
  +\frac{1}{4}\left(\ro\nu''(\ro)+\nu'(\ro)\right)\left\{2\difk + H_1\cos\vartheta\right\}\nonumber\\
  &&\qquad-\frac{1}{4}(H_1+2c_1)\nu'(\ro)e^{\lambda(\ro)}\cos\vartheta. \label{app:difhp}
\end{eqnarray}
Finally,
although equation $[\kappaperper_{\vartheta\phi}]=0$ is automatically satisfied,
in this case the equation $[\kappaperper_{\tau\phi}]=0$ provides the condition
$
[\omega'']Q_1=0.
$

As in the first order case, the constants
$D_1$ and $D_2$  
can be safely absorbed by using a isomorphic \emph{spacetime} gauge at one (any) side,
say $\vec V_2^+=D_1\partial_{\tp}+D_2\partial_{\phip}$,
keeping $\vec s_1=0$. Clearly $\vec s_2^+=\vec V_2^+$ and therefore
by (\ref{eq:Z_gauges}) that leads to $\vec T^+_2\to \vec T^+_2-\vec s_2^+$ and
the second order metric perturbation tensor
$\spt^+$ is unchanged. We can thus set $D_1=D_2=0$ without loss of generality.\fin

\section*{Acknowledgements}
We thank Marc Mars for his most valuable suggestions and discussions.
We are also grateful to Michael Bradley and Alfred Molina for their comments and suggestions. We finally also thank a referee for his corrections, observations and suggestions.
We acknowledge financial support from projects IT592-13 (GIC12/66)
of the Basque Government, FIS2010-15492 from the MICINN, and
UPV/EHU under program UFI 11/55.
BR acknowledges financial support from the Basque Government grant BFI-2011-250.

The full calculations have been performed using computer algebra systems,
in particular, the free PSL version of REDUCE.

\bibliography{references}{}

\begin{thebibliography}{10}

\bibitem{Battye01}
R.~A. Battye and B.~Carter (2001) \emph{Generic junction conditions in
  brane-world scenarios}.
\newblock \PLB \textbf{509} 331.

\bibitem{Bonnor_Vickers}
W.~B. Bonnor and P.~A. Vickers (1981) \emph{Junction conditions in {G}eneral
  {R}elativity}.
\newblock \GRG \textbf{13} 29--36.

\bibitem{Bradley_etal2007}
M.~Bradley, D.~Eriksson, G.~Fodor and I.~R\'acz (2007) \emph{Slowly rotating
  fluid balls of {P}etrov type {D}}.
\newblock \PRD \textbf{75} 024013.

\bibitem{Brizuela2010}
D.~Brizuela, J.~M. Martín-García, U.~Sperhake and K.~D. Kokkotas (2010)
  \emph{High-order perturbations of a spherical collapsing star}.
\newblock Phys. Rev. D \textbf{82} 104039.

\bibitem{Bruni_et_al_1997}
M.~Bruni, S.~Matarrese, S.~Mollerach and S.~Sonego (1997) \emph{Perturbations
  of spacetime: gauge transformations and gauge invariance at second order and
  beyond}.
\newblock \CQG \textbf{14} 2585--2606.

\bibitem{CMMR}
J.~Cabezas, J.~Mart\'in, A.~Molina and E.~Ruiz (2007) \emph{An approximate
  global solution of {E}instein’s equations for a rotating finite body}.
\newblock \GRG \textbf{39} 707--736.

\bibitem{Chandra_Poly_Newton_1933}
S.~Chandrasekhar (1933) \emph{The equilibrium of distorted polytropes. {I}.
  {T}he rotational problem}.
\newblock \MNRAS \textbf{93} 390--406.

\bibitem{ChandraLebo_Newton_1962}
S.~Chandrasekhar and N.~R. Lebovitz (1962) \emph{On the oscillations and the
  stability of rotating gaseous masses.}
\newblock \APJ \textbf{135} 248.

\bibitem{Chandra_Miller1974}
S.~Chandrasekhar and J.~C. Miller (1974) \emph{On slowly rotating homogeneous
  masses in {G}eneral {R}elativity}.
\newblock \MNRAS \textbf{167} 63--80.

\bibitem{ColpiMiller}
M.~Colpi and J.~C. Miller (1992) \emph{Rotational properties of strange stars}.
\newblock \APJ \textbf{388} 513--520.

\bibitem{CuchiLinear}
J.~E. Cuchí, A.~Gil-Rivero, A.~Molina and E.~Ruiz (2013) \emph{An approximate
  global solution of {E}instein’s equation for a rotating compact source with
  linear equation of state}.
\newblock \GRG \textbf{45} 1457.

\bibitem{Tesis_D_Eriksson}
D.~Eriksson (2008) \emph{Perturbative Methods in {G}eneral {R}elativity}.
\newblock Ph.D. thesis, Ume\aa \, Universitet.

\bibitem{Gonzalez_Romero_etal}
L.~M. Gonz\'alez-Romero and J.~L. Bl\'azquez-Salcedo (2009) \emph{Core-crust
  transition pressure evolution in post-glitch epoch for a vela-type pulsar}.
\newblock Arxiv:0912.0628.

\bibitem{Hartle1967}
J.~B. Hartle (1967) \emph{Slowly rotating relativistic stars. {I}. {E}quations
  of structure}.
\newblock \APJ \textbf{150} 1005--1029.

\bibitem{HartleThorne1968}
J.~B. Hartle and K.~S. Thorne (1968) \emph{Slowly rotating relativistic stars.
  {II}. {M}odels for neutron stars and supermassive stars}.
\newblock \APJ \textbf{153} 807--834.

\bibitem{HartleThorne1969}
J.~B. Hartle and K.~S. Thorne (1969) \emph{Slowly rotating relativistic stars.
  {III}. {S}tatic criterion for stability}.
\newblock \APJ \textbf{158} 719--726.

\bibitem{Lattimer2012}
J.~M. Lattimer (2012) \emph{The nuclear equation of state and neutron star
  masses}.
\newblock Ann. Rev. Nucl. Part. Sci. \textbf{62} 485--515.

\bibitem{MaMaVe2007}
M.~A.~H. MacCallum, M.~Mars and R.~Vera (2007) \emph{Stationary axisymmetric
  exteriors for perturbations of isolated bodies in {G}eneral {R}elativity, to
  second order}.
\newblock \PRD \textbf{75} 024017.

\bibitem{Mars2005}
M.~Mars (2005) \emph{First- and second-order perturbations of hypersurfaces}.
\newblock \CQG \textbf{22} 3325--3348.

\bibitem{Mars2007}
M.~Mars, F.~C. Mena and R.~Vera (2007) \emph{Linear perturbations of matched
  spacetimes: the gauge problem and background symmetries}.
\newblock \CQG \textbf{24} 3673--3689.

\bibitem{Mars99}
M.~Mars and J.~M.~M. Senovilla (1999) \emph{Geometry of general hypersurfaces
  in spacetime: junction conditions}.
\newblock Classical and Quantum Gravity \textbf{10} 1865.

\bibitem{CMMRpolitropo}
J.~Mart\'in, A.~Molina and E.~Ruiz (2008) \emph{Can rigidly rotating polytropes
  be sources of the {K}err metric?}
\newblock \CQG \textbf{25} 105019.

\bibitem{JMMG-Gundlach}
J.~Mart\'in-Garc\'ia and C.~Gundlach (2001) \emph{Gauge-invariant and
  coordinate-independent perturbations of stellar collapse. ii. matching to the
  exterior}.
\newblock \PRD \textbf{64} 024012.

\bibitem{Mukohyama00}
S.~Mukohyama (2000) \emph{Gauge-invariant gravitational perturbations of
  maximally symmetric spacetimes}.
\newblock \PRD \textbf{62} 084015.

\bibitem{Borja_constant_rho}
B.~Reina (2015) \emph{{Slowly rotating homogeneous masses revisited}}.
\newblock {arXiv:1503.07835} .

\bibitem{ReinaVera_ERE2012}
B.~Reina and R.~Vera, \emph{Revisiting {H}artle's model for relativistic
  rotating stars}.
\newblock In A.~Garc\'ia-Parrado, F.~C. Mena, F.~Moura and E.~Vaz, eds.,
  \emph{Progress in Mathematical Relativity, Gravitation and Cosmology},
  vol.~60 (Springer, Heideberg, New York, Dordrecht, London 2014), pages
  377--381.

\bibitem{Exact_solutions}
H.~Stephani, D.~Kramer, M.~MacCallum, C.~Hoenselaers and E.~Herlt, \emph{Exact
  Solutions of Einstein’s Field Equations} (Cambridge University Press 2003),
  second edn.
\newblock Cambridge Books Online.

\bibitem{LivRev_Stergioulas}
N.~Stergioulas (2003) \emph{Rotating stars in {R}elativity}.
\newblock Living Reviews in Relativity \textbf{6}.

\bibitem{Vera2002}
R.~Vera (2002) \emph{Symmetry-preserving matchings}.
\newblock \CQG \textbf{19} 5249--5264.

\end{thebibliography}
\bibliographystyle{review_bib}

\end{document}